\documentclass[manuscript,screen]{acmart}
\AtBeginDocument{%
  }

\setcopyright{acmlicensed}
\copyrightyear{2018}
\acmYear{2018}
\acmDOI{XXXXXXX.XXXXXXX}
\acmConference[Conference acronym 'XX]{Make sure to enter the correct
  conference title from your rights confirmation email}{June 03--05,
  2018}{Woodstock, NY}
\acmISBN{978-1-4503-XXXX-X/2018/06}

\usepackage{booktabs} 
\usepackage{array}    
\usepackage{tabularx} 
\usepackage{multirow} 
\usepackage{subcaption}
\usepackage[ruled,vlined]{algorithm2e}
\usepackage[commandnameprefix=always]{changes} 

\usepackage[svgnames]{xcolor} 
\setdeletedmarkup{\colorbox{red!30}{\sout{#1}}} 
\setaddedmarkup{\colorbox{cyan!30}{#1}}      

\begin{document}

\title[Co-Authoring the Self]{Co-Authoring the Self: A Human-AI Interface for Interest Reflection in Recommenders}

\author{Ruixuan Sun}
\affiliation{%
 \institution{Grouplens Research, University of Minnesota}
 \streetaddress{5-244 Keller Hall, 200 Union Street SE}
 \city{Minneapolis}
 \state{Minnesota}
 \country{United States}}

 \author{Junyuan Wang}
\affiliation{%
 \institution{Department of Computer Science and Engineering, University of Minnesota}
 \city{Minneapolis}
 \state{Minnesota}
 \country{United States}}

 \author{Sanjali Roy}
\affiliation{%
 \institution{Department of Computer Science and Engineering, University of Minnesota}
 \city{Minneapolis}
 \state{Minnesota}
 \country{United States}}

 \author{Joseph A. Konstan}
 \affiliation{%
 \institution{Grouplens Research, University of Minnesota}
 \streetaddress{5-244 Keller Hall, 200 Union Street SE}
 \city{Minneapolis}
 \state{Minnesota}
 \country{United States}}

\renewcommand{\shortauthors}{Trovato et al.}

\begin{abstract}
    Natural language–based user profiles in recommender systems have been explored for their interpretability and potential to help users scrutinize and refine their interests, thereby improving recommendation quality. Building on this foundation, we introduce a human–AI collaborative profile for a movie recommender system that presents editable personalized interest summaries of a user’s movie history. Unlike static profiles, this design invites users to directly inspect, modify, and reflect on the system’s inferences. In an eight-week online field deployment with 1775 active movie recommender users, we find persistent gaps between user-perceived and system-inferred interests, show how the profile encourages engagement and reflection, and identify design directions for leveraging imperfect AI-powered user profiles to stimulate more user intervention and build more transparent and trustworthy recommender experiences.
\end{abstract}



\maketitle

\section{Introduction}

User profiles are a key element in personalized recommendation systems, yet many existing implementations lack interactivity, transparency, and granularity, often reducing them to coarse statistics of user clicks or high-level metadata summaries \cite{10.1145/3477495.3531873, bang2025llmbaseduserprofilemanagement, 10.1145/3331184.3331211}. An up-to-date and accurate profile is critical for producing high-quality recommendations, but current approaches face limitations. Traditional rating- or item-based profiles may fail to capture nuanced preferences—such as thematic interests, emotional resonance, or contextual dislikes—that go beyond numeric scores \cite{chen2015recommender}.

This is an area where AI, particularly generative AI (Gen-AI), can help. Articulating preferences explicitly is often difficult for users, but Gen-AI can scaffold the process by helping them articulate their tastes, recognize misalignments in system-generated descriptions, and provide more effective critiques. Research in human–AI collaboration (HAC) suggests that AI should not only assist in task completion but also help users understand and manage their goals \cite{10.1145/3334480.3381069}. While HAC has shown promise in improving productivity and recommendation quality, it also raises concerns about reduced user agency and ownership \cite{10.1145/3331184.3331211, 10.1145/3613904.3642134, 10.1145/3677081}. In recommender systems specifically, HAC has the potential to improve transparency and trust by exposing how the system perceives a user’s preferences \cite{knijnenburg2012explaining, pu2012evaluating, 10.1145/3477495.3531873, tintarev2010designing}.

Despite these advances, few natural language–based interfaces support both understanding and exploration of user interests in recommender systems. To address this gap, we adopt a human–AI collaborative approach to design a novel user profile page on the live movie recommendation platform MovieLens (\url{https://movielens.org/})\footnote{We appreciate the support and assistance from the MovieLens team.}. Our work introduces a large language model (LLM)-powered “self-portrait” interface, co-created by humans and AI, that dynamically generates textual summaries of a user’s movie interests based on historical ratings. The interface supports user editing and regenerates summaries when new ratings are added, enabling continuous interest representation and user reflection.


This three-phase study investigates how to build an AI-powered critiquing-based user profile interface in recommender systems for: (1) understanding user needs and the limitations of the current rating profile; (2) designing an interactive, LLM-augmented interest interface; and (3) evaluating its short-term utility and long-term behavioral effects. Specifically, we address the following research questions:

\begin{quote}
    \emph{\textbf{RQ1} (Limitations \& Goals)}: How do users perceive the limitations of current profile interfaces, and what goals do they have for improving their ability to express and maintain their preferences?
    
    \emph{\textbf{RQ2} (Enhanced Awareness \& Understanding)}: To what extent does the new human–AI collaborative profile interface enhance users’ awareness and understanding of their interests?  
    
    \emph{\textbf{RQ3} (Long-term Interaction Patterns)}: How does the interactive, co-reflective design of the interface influence users’ long-term interaction patterns with the recommender system?
\end{quote}

The remainder of this paper first reviews relevant literature, then presents our study methods and the design of the new user profile interface. We next describe the online field experiment and evaluation results from an eight-week real-user trial. Finally, we conclude with discussion and directions for future work.

\section{Related Work}

\subsection{LLMs for Human-AI Collaboration}

The growing capabilities of Gen-AI tools, especially LLMs, have sparked significant research interest in human–AI collaboration (HAC). \citeauthor{10.1145/3334480.3381069} argue that effective HAC requires a shift toward AI systems that support mutual understanding and co-management of goals, moving beyond the traditional view of AI as a task-execution tool \cite{10.1145/3334480.3381069}. However, the impact of LLMs on collaborative dynamics is complex. For example, \citeauthor{10.1145/3677081} found that while an LLM assistant improved cognitive performance in a cooperative game, it also created challenges for team understanding \cite{10.1145/3677081}.

Interaction design is a critical factor in successful HAC. \citeauthor{10.1145/3613905.3651042} demonstrated how varying levels of agency in LLM-aided data analysis tools influence user perceptions of LLM outputs \cite{10.1145/3613905.3651042}. Similarly, \citeauthor{10.1145/3613904.3642134} examined AI scaffolding in co-writing, finding that while increased AI support improved writing productivity, it sometimes reduced users’ sense of ownership over the resulting text \cite{10.1145/3613904.3642134}.

Applying HAC to personal reflection tasks introduces unique considerations. For example, \citeauthor{10.1145/3613904.3642693} studied LLM-assisted journaling and found that while it encouraged new perspectives in user reflections, it also raised concerns about over-reliance on AI-generated emotional content \cite{10.1145/3613904.3642693}. Other studies have examined HAC in annotation tasks \cite{kim-etal-2024-meganno} and remote sighted assistance \cite{fi16070254}. Our work focuses on leveraging LLM collaboration for the reflective process of interacting with user profiles in recommender systems, with the aim of enhancing both user agency and awareness of personal interests through co-created artifacts.

\subsection{Reflective Interfaces and Exploration in Recommender Systems}

Recent literature in recommender systems have moved beyond purely algorithmic concerns to the broader user experience, emphasizing the role of interfaces in shaping perceptions, behavior, and long-term engagement \cite{lu2023user, jannach2021survey, he2023survey, chen2023bias, maslowska2022role, xue2023prefrec}. A growing focus is on reflective interfaces that are designed to help users become more aware of their preferences and interaction patterns. \citeauthor{eiband2021support} argue that intelligent systems like recommender systems should be designed for reflection, enabling users to better understand their behaviors and values through interaction with the system \cite{eiband2021support}. Similarly, \citeauthor{10.1145/3597499} presented a critiquing-based conversational recommender that allows users to specify directional preferences on item attributes; they found this approach enhanced both recommendation quality and overall satisfaction \cite{10.1145/3597499}. These works suggest that interfaces that reveal, rather than merely explain, can also enrich user experiences.

Furthermore, prior arts on profile transparency reinforces the importance of reflective interfaces. Studies show that exposing users to the system’s understanding of their preferences can foster trust, perceived control, and satisfaction \cite{tintarev2010designing, knijnenburg2012explaining, pu2012evaluating}. While most explanations in recommender systems are item-based—focusing on why a specific recommendation is made, recent work also examines how system-generated representations of user profiles can function as reflective artifacts \cite{tsai2021effects}.


\subsection{User Interest Summarization}

Recent advances in recommender systems show increasing use of LLMs to generate natural language user profiles. According to \citeauthor{10.1145/3477495.3531873}, such profiles that summarize user preferences can offer several advantages, including greater transparency, opportunities for users to review and modify inferred preferences, and adaptability to significant preference shifts \cite{10.1145/3477495.3531873}. Different approaches exist for creating natural language profiles. Tag-based methods cluster keywords from user-rated items and assign semantic labels. \citeauthor{Yang_Song_Ji_2015} found such tags to outperform traditional content-based methods for interest inference \cite{Yang_Song_Ji_2015}. Modern implementations often use BERTopic, a BERT-powered topic model \cite{grootendorst2022bertopicneuraltopicmodeling, Devlin_Chang_Lee_Toutanova_2019}, as in \citeauthor{Hill_Goo_Agarwal_2025}’s university-program recommender, which achieved 98\% alignment with users’ self-reported interests \cite{Hill_Goo_Agarwal_2025}. Non-tag methods include set-based systems, where tags describe item sets rather than clusters, addressing labeling challenges \cite{10.1145/3331184.3331211}. Others generate profiles from live behavior (e.g., website visits \cite{10.1145/3701571.3703373}) or use LLMs to enhance profiles by integrating review text and purchase data \cite{bang2025llmbaseduserprofilemanagement}. \citeauthor{10.1145/2433396.2433492} further improved profile accuracy by prioritizing key user interactions and using hybrid algorithms \cite{10.1145/2433396.2433492}.

\section{Method}

We adopted a mixed-methods design consisting of three stages, which was reviewed and approved by our institutional review board (IRB).

\begin{enumerate}
    \item Distributing \textit{formative survey} to understand active users’ sentiment toward the default static profile interface and their expectations for new interactive features.
    \item Conducting \textit{online field experiment} to evaluate a new interactive profile interface incorporating a personalized movie interest summary, co-edited asynchronously by the user and a generative AI tool.
    \item Running \textit{log and post-survey analysis} to examine human--recommender interactions through system logs and post-experiment subjective feedback.
\end{enumerate}

\subsection{Formative Survey}

\begin{figure}[!htbp]
\centering
  \includegraphics[width=0.45\columnwidth]{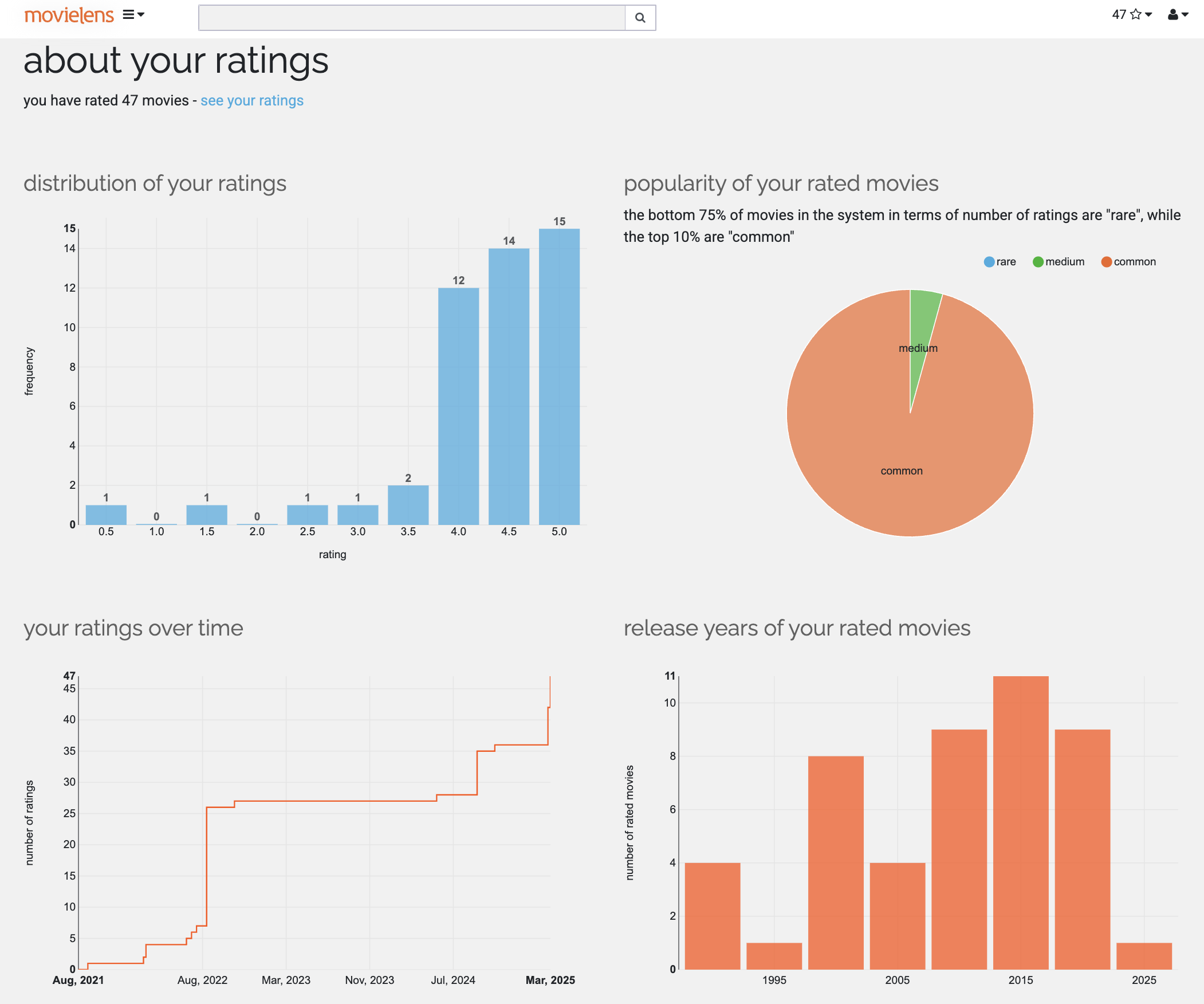}
  \includegraphics[width=0.45\columnwidth]{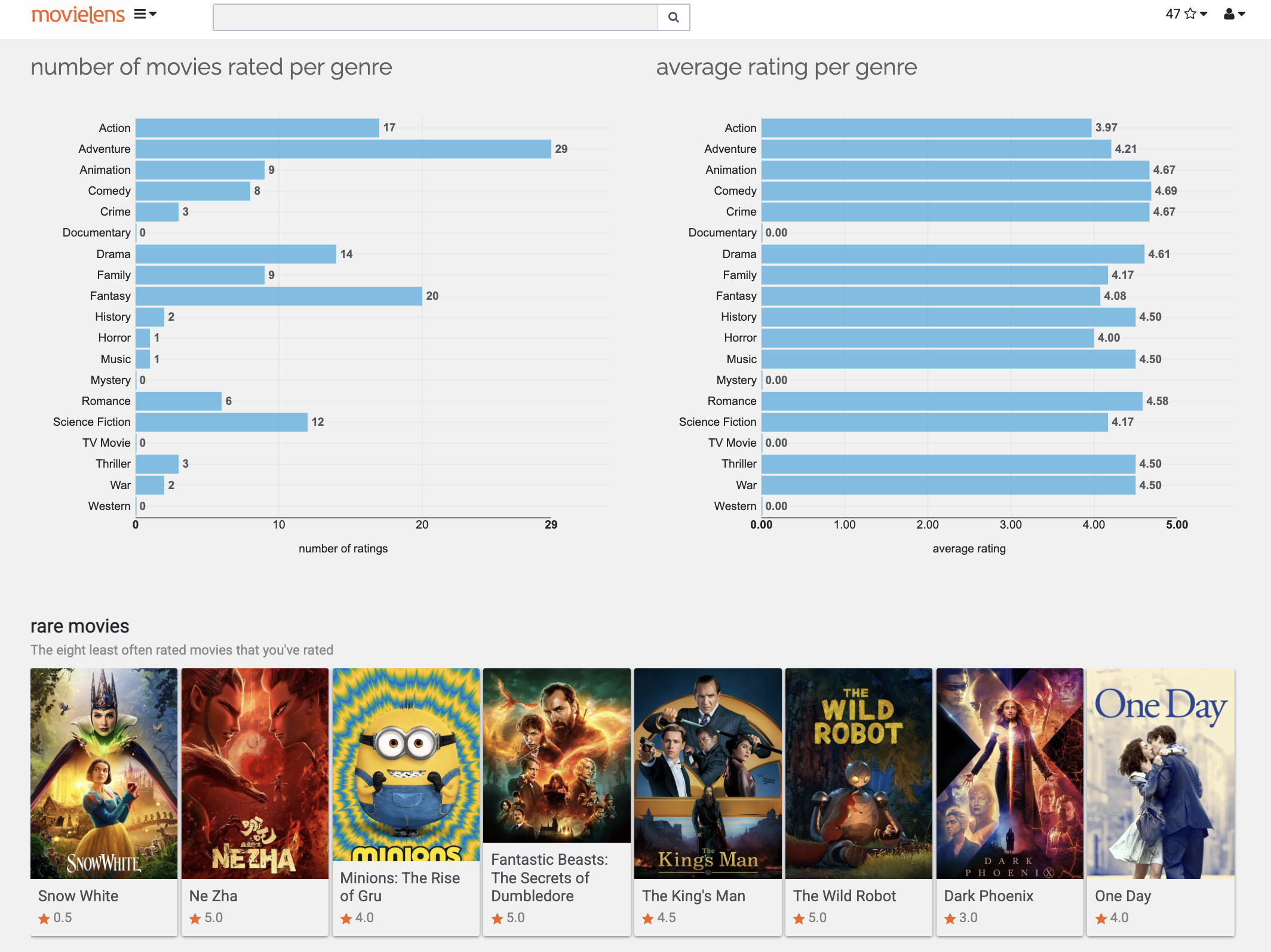}
  \caption{The default MovieLens "About your ratings" page. We asked users to evaluate their experience with the interface during the formative survey. }~\label{fig:default-ml-page}
\end{figure}

The formative survey included five Likert-style questions about the pre-existing MovieLens user rating profile page (Fig.~\ref{fig:default-ml-page}), covering:  
\begin{itemize}
    \item The usefulness of the profile for understanding one’s preferences.
    \item The ease of interpreting individual movie recommendations.
    \item Overall satisfaction with the profile experience.
\end{itemize}

In addition, participants wrote a short paragraph describing their taste in movies and their expectations for a new profile interface. The full set of survey questions is listed in Appendix Table~\ref{tab:pre_survey_questions}.

We distributed the survey on October 23, 2024, to 2,122 active MovieLens users (defined as having logged in more than six times and rated more than ten movies between April and October 2024). We received 349 responses within two weeks, of which 266 were complete and used for analysis.

\begin{quote}
    \textbf{RQ1}:\emph{How do users perceive the limitations of current profile interfaces, and what goals do they have for improving their ability to express or maintain their preferences?}
\end{quote}

We analyzed both quantitative ratings and qualitative open-ended responses. As shown in the left plot of Fig.~\ref{fig:pre-post-survey-stats}, users generally found the default profile helpful for understanding preferences, but less effective for understanding recommendations. We then applied grounded theory thematic analysis to qualitative responses describing desired new features, and identified two primary themes: 

\begin{enumerate}
    \item \textbf{Explain interests} --- Provide AI-generated summaries of user tastes and clarify why movies are recommended, possibly by comparing them to previously enjoyed films.
    \item \textbf{Visualized exploration tools} --- Offer tools to inspect favorite movie metadata, such as actor or director frequency, or the distribution of languages, origins, and popularity among rated films.
\end{enumerate}

Further coding revealed a strong desire for transparency in recommendation rationales. Many participants explicitly requested concise, AI-generated textual explanations describing their preferences and showing how these preferences influenced recommendations. Several suggested short comparisons between recommended movies and those previously enjoyed, while others proposed displaying concrete factors such as genre, director, or actors. These insights directly informed the design of the new interactive profile interface, which is covered in more detail in Sec. \ref{new-interface-section}.

\subsection{Online Field Experiment}

After designing the new user profile interface, we evaluate the effectiveness and user reaction to it in a naturalistic setting and conducted a longitudinal online field experiment with active users\footnote{Active users are defined as those who logged in over 12 times and rated over 20 movies in the year of 2024.} on the MovieLens website. The trial ran for eight weeks, from March~29,~2025 to May~24,~2025. During this period, 5,804 users visited the website, of whom 1,775 met the active user criteria and were thus qualified participants. The experimental interface was made available through a ``Your Interest Summary'' button placed prominently on each participant’s home page. In addition, users were prompted with a pop-up message upon their first login during the study period to inform them of the new feature. They were also informed that clicking on the button to access the new interface and editing it was completely voluntary.

\subsection{Log Analysis}
\label{sec:log-analysis}

To investigate the indirect effects of the new interest-summary interface on longitudinal user--recommender interaction, we drew on metrics from prior work~\cite{zhao2018explicit,knijnenburg2012explaining} and categorized behavioral measures into four dimensions: \textbf{Exploration}, \textbf{Engagement}, \textbf{Reflection}, and \textbf{Satisfaction}. Detailed metric definitions are provided in Table~\ref{tab:log-analysis}. For diversity-related measures of viewed and rated movies, we computed \emph{intra-list similarity} (ILS)~\cite{jesse2023intra} using each movie’s tag embeddings~\cite{vig2011navigating}.

\subsection{Post-Experiment Survey}

After the experiment, we invited participants who had edited their interest profile on the new interface at least once ($N = 216$) to complete a post-experiment survey. In addition to general usability and satisfaction questions, the survey collected feedback on each of the three interest-summary text boxes for their helpfulness for understanding movie preferences. The complete set of questions is provided in Appendix~\ref{tab:post_survey_questions}. Fifty-seven users clicked the survey link, and 26 completed it while confirming that they had used the interface (by answering ``Yes'' to Q1). We only used the completed response data for analysis.

\begin{table}[h]
\centering
\begin{tabularx}{\linewidth}{l>{\ttfamily}lX}
\toprule
\textrm{Category} & \textrm{Metric Name} & \textrm{Description} \\
\midrule
\multirow{4}{*}{Engagement} 
    & movie\_view\_count & Number of movies a user viewed during experiment \\
    & rating\_count & Number of movies a user rated during experiment \\
    & login\_count & Number of unique sessions a user logs in during experiment \\
    & session\_length & Total hours of a user's dwelling time with the recommender during experiment \\
\cmidrule(lr){1-3}
\multirow{2}{*}{Exploration} 
    & rated\_movie\_div & Diversity of a user's rated movies based on their tag embeddings range in [0,1], defined by intra-list similarity: ${\frac{N(N-1)}{2}}\sum_{i=1}^{N-1} \sum_{j=i+1}^{N} \rho(\vec{v_i}, \vec{v_j})$ \cite{jesse2023intra} \\
    & viewed\_movie\_div & Diversity of a user's viewed movies based on their tag embeddings range in [0,1], same definition as ILS eqution above. \\
\cmidrule(lr){1-3}
Reflection 
    & rerate\_total & Number of movies a user re-rated during experiment \\
\cmidrule(lr){1-3}
Satisfaction 
    & avg\_rating & Average score of movies user rated during experiment, each rating can range from [0.5, 5] in half-point step \\
\bottomrule
\end{tabularx}
\caption{Log analysis metrics.}
\label{tab:log-analysis}
\end{table}

\section{New Interactive User Interest Profile} \label{new-interface-section}

\begin{figure}[!htbp]
\centering
  \includegraphics[width=0.8\columnwidth]{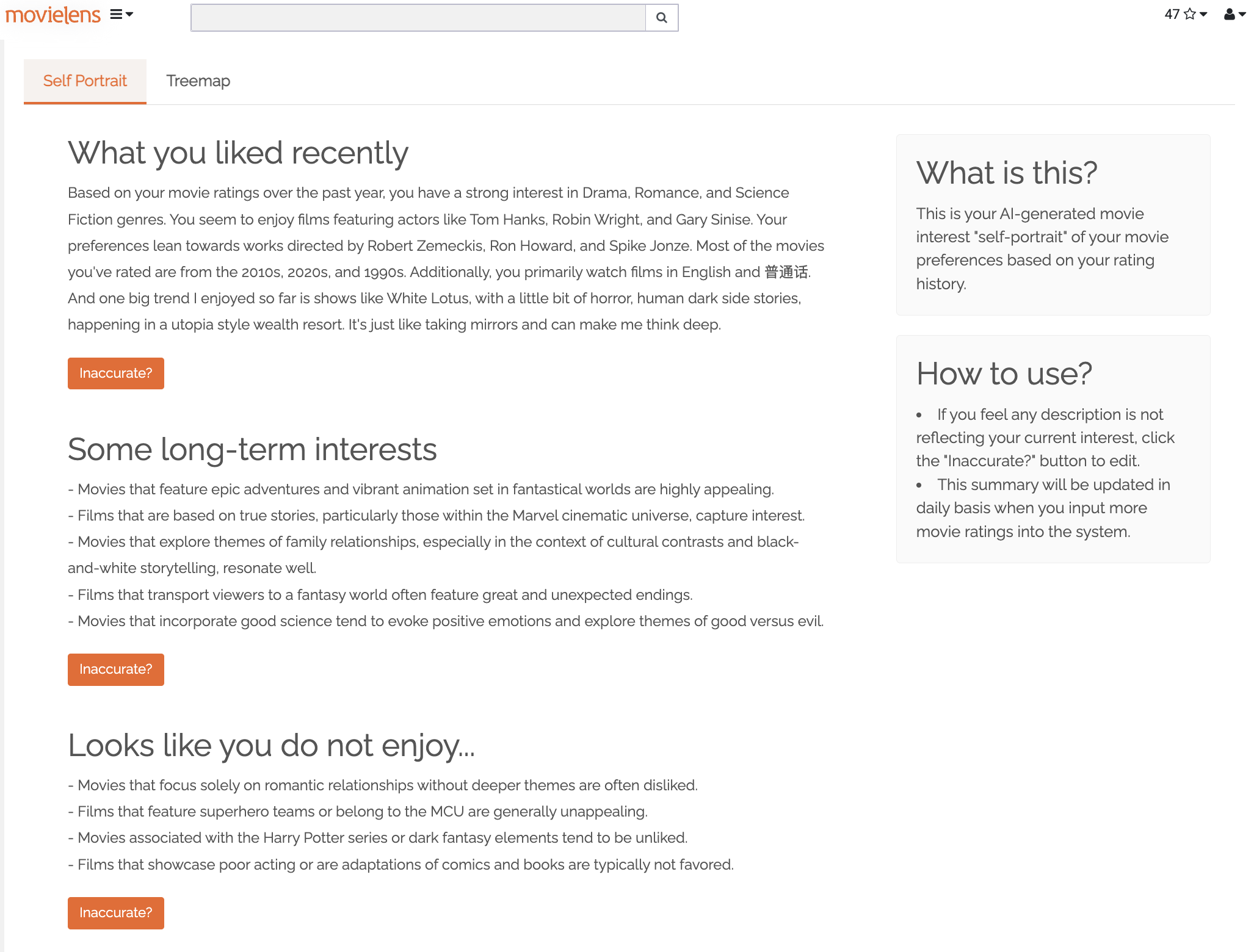}
  \includegraphics[width=0.45\columnwidth]{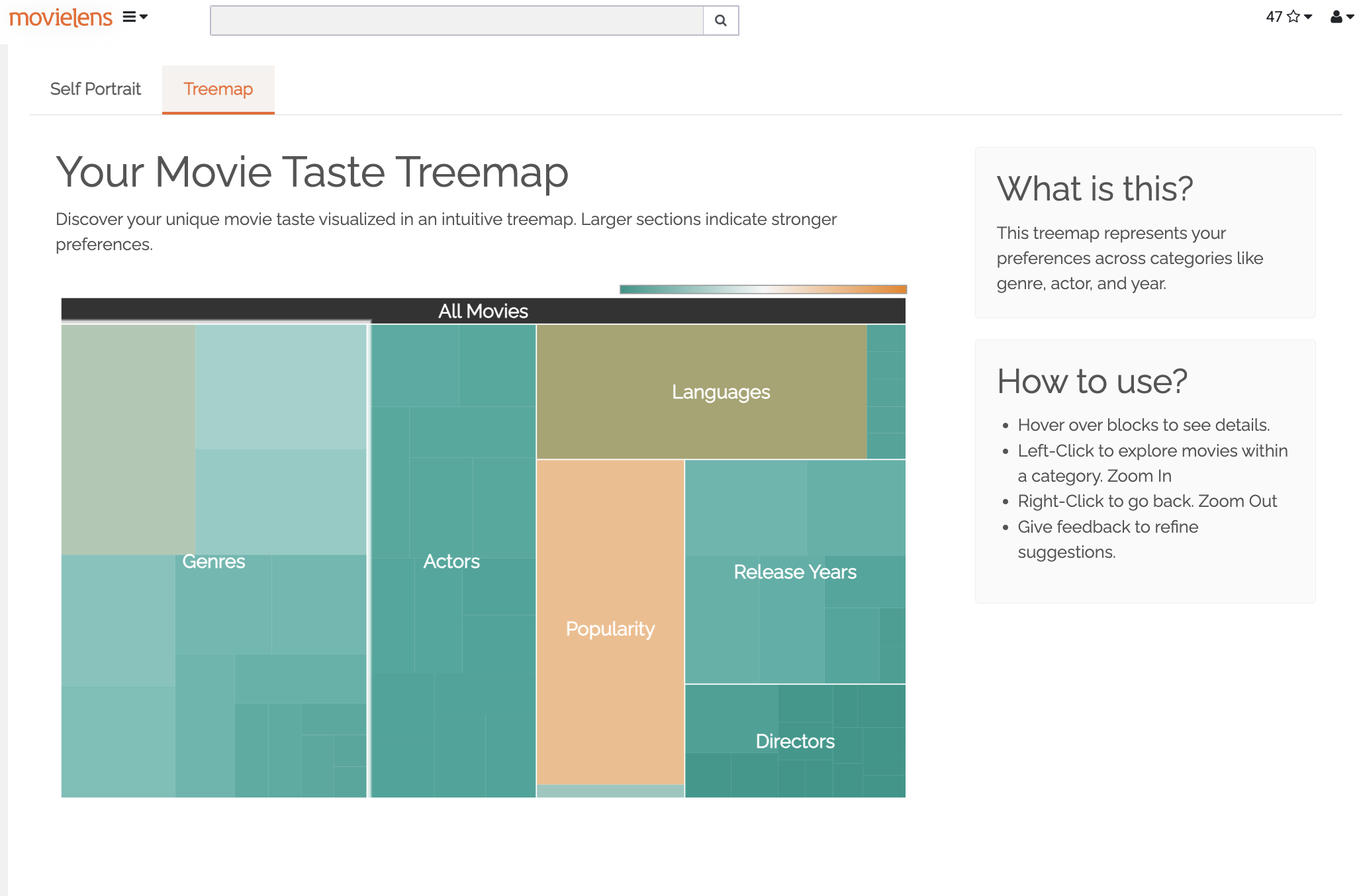}
  \includegraphics[width=0.45\columnwidth]{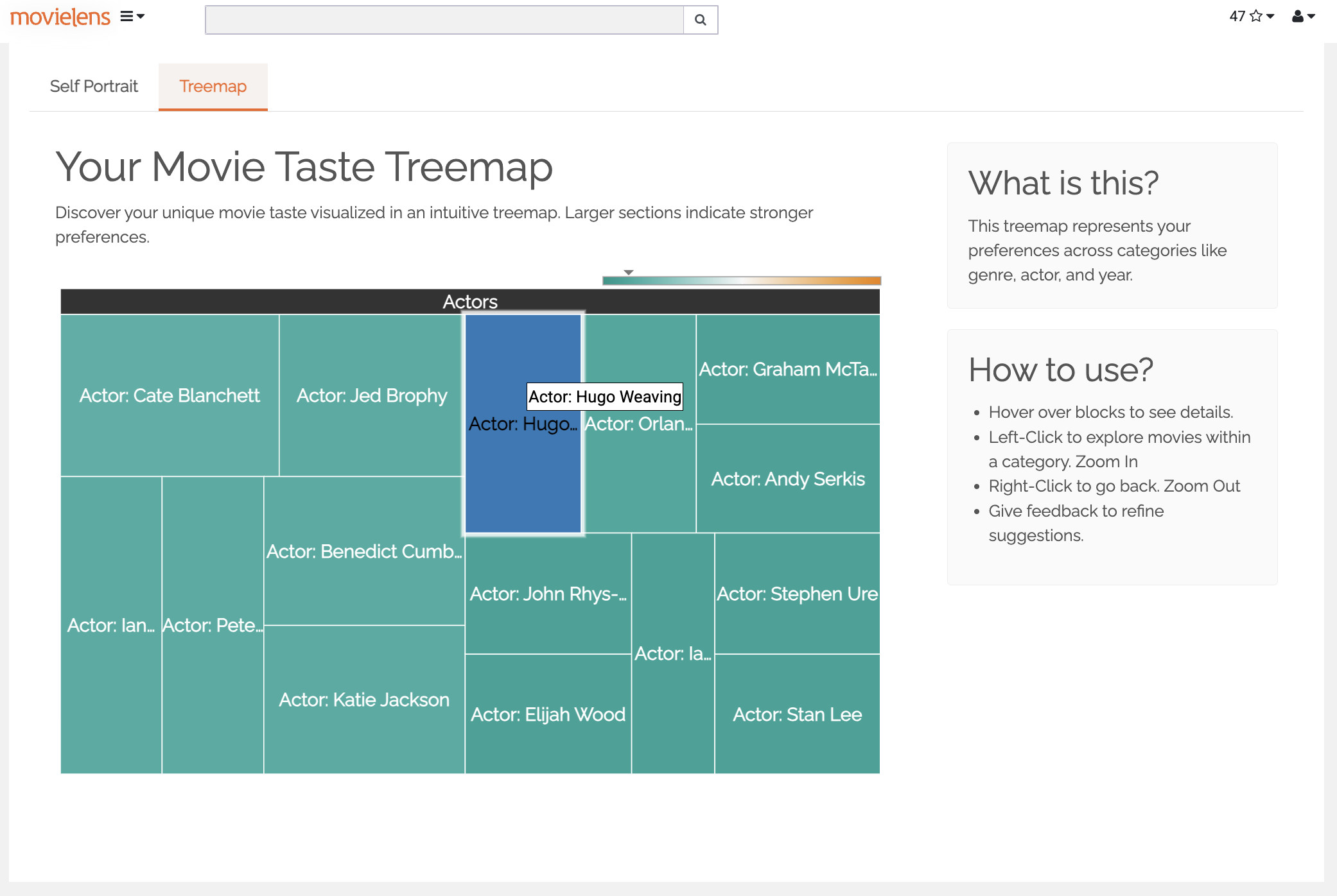}
  \caption{Example interface for the movie interest self-portrait and categorical treemap visualization. Users can toggle between the two views using the tab button in the top-left corner. \textbf{Top:} The self-portrait interface, allowing users to view, edit, and save their current interest summaries. \textbf{Bottom:} The movie taste treemap that users could use to explore their rated movies in six categories, with zoom-in and zoom-out interactions available for each block.}
  \label{fig:interface}
\end{figure}

Based on findings from our formative survey, we designed a new interactive profile centered on a "self-portrait" of the user’s movie tastes. This primary component comprises three editable text fields summarizing the user’s recent (one-year) interests, long-term enjoyed movies, and long-term disliked movies. It is also part of a larger interface including a treemap visualization that user can use to interactively explore the distribution of their rated movies across six categories (Genre, Actor, Director, Language, Popularity, Release Year). Note our paper does not look for use of statistics for treemap and it is not the focus of the study. An overview of the new interface is shown in Fig.~\ref{fig:interface}.

\subsection{AI-Powered Summary Generation and Refinement}

\begin{figure}[!htbp]
\centering
  \includegraphics[width=1.0\columnwidth]{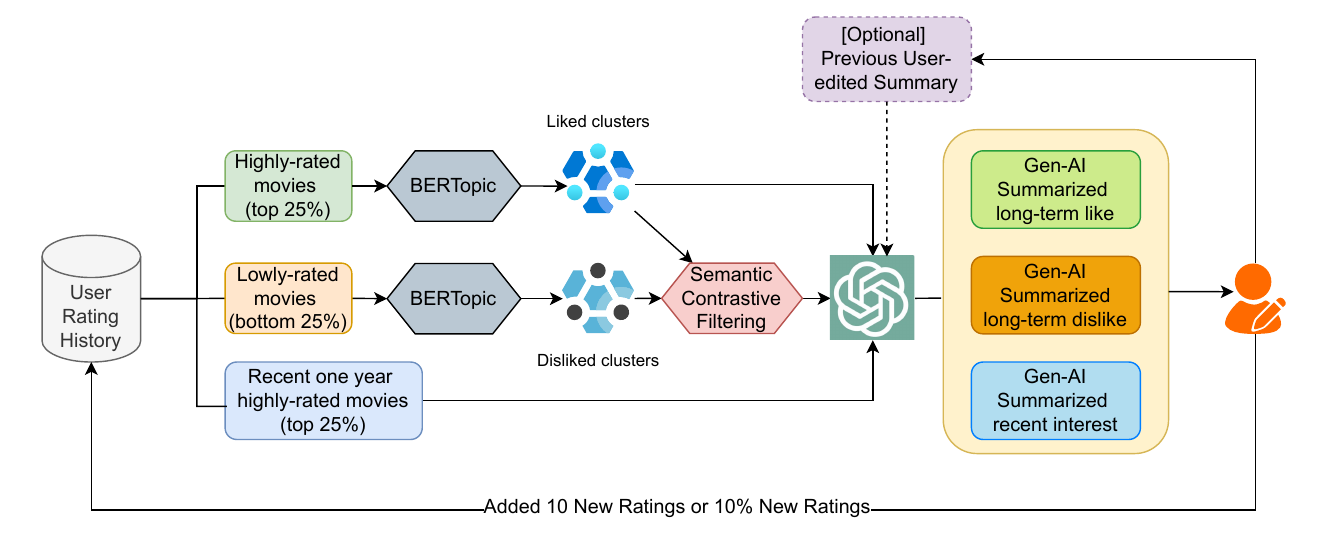}
  \caption{User interest summary generation pipeline. The dotted box indicates the optional user-edited summary (not available in initial session), which is incorporated as context in subsequent regeneration. The pipeline runs for each qualified user at experiment start, and loops after rating activity exceeds the defined threshold (added 10\% of total ratings or 10 more new ratings after each generation).}
  \label{fig:interest-generation-pipeline}
\end{figure}

The textual summaries for the self-portrait were generated and maintained through a systematic pipeline, detailed in Fig.~\ref{fig:interest-generation-pipeline}. 
First, we extracted three sets of movies for each user from the MovieLens database: historically highly-rated (top 25\%), historically low-rated (bottom 25\%), and recent one-year highly-rated (top 25\%). To validate the meaningful differences on two rating quartiles, we plotted the max-bottom and min-top rating cut-off on the highly- and lowely-rated movies for each user and demonstrate a significant and meaningful gap for the majority of users (see Appendix Fig. \ref{fig:top-bottom-rating-quartiles}). We then generated personalized summaries for these sets using two distinct strategies. Example prompts can be found in Appendix \ref{fig:generation_prompt}.

\begin{itemize}
    \item \textbf{Long-term enjoyed and disliked summaries:} To synthesize interests from a large number of historic ratings, we leveraged community-contributed movie tags. As detailed in Algorithm~\ref{algo:1}, we clustered the top-10 community tags for each movie in a user's rating history using a BERTopic model to identify up to five distinct semantic interest groups. A LangChain DataFrame agent, powered by the GPT-4o-mini model, then summarized each cluster into a single sentence. To avoid contradictions (e.g., liking one film in a series but disliking another, even if they share similar top-10 tags), we applied contrastive filtering, removing any "disliked" cluster that shared high semantic similarity ($cos(\theta) \geq 0.8$) with a "liked" cluster.
    
    \item \textbf{Short-term recent interest summary:} To capture more specific, movie-level details from a user's recent activity, we prompted the same AI agent to produce a five-sentence summary based on the top genres, actors, directors, release years, and languages from the user's highly-rated movies in the past year.
    
\end{itemize}

\begin{algorithm}[t]
    \caption{Long-term User Interest Extraction and Filtering}
    \label{algo:1}
    \KwIn{User $u$; Extract rated movies $M_u = \{m_1, m_2, \dots, m_n\}$}
    \KwOut{Refined summary sentences for liked and disliked movie interest}
    \ForEach{movie $m \in M_u$}{
        Collect top-10 community tags $T(m)$\;
    }
    Cluster $\bigcup_{m \in M_u} T(m)$ into up to 5 groups using BERTopic\;
    \ForEach{disliked cluster $c^-$}{
        \ForEach{liked cluster $c^+$}{
            \If{$\cos(\mathbf{e}_{c^-}, \mathbf{e}_{c^+}) > 0.8$}{
                Remove $c^-$ (contrastive filtering)\;
            }
        }
    }
    \ForEach{cluster $c$}{
        Summarize $c$ into one sentence using Gen-AI Agent\;
    }
\end{algorithm}

To ensure the reliability of this process, we evaluated the faithfulness of the generated summaries. Three researchers conducted subjective assessment and manually compared a random sample of 100 summary triplets against their underlying BERTopic clusters, finding that 98\% of summaries were factually aligned with the source topics.

Finally, the interface was designed for continuous interaction. After the initial generation, users could edit and refine their summaries at any time. At the backend, the interest refinement system would also automatically trigger a regeneration of the summaries to incorporate new preferences. This occurred daily if a user's rating activity exceeded a set threshold (adding either 10\% of their previous total ratings or 10 new ratings). During regeneration, any user-edited text was fed back into the prompt as additional context, enabling a co-creative loop between the user and the AI. Note that the regular recommender pipeline is not impacted by this new interface loop.

\section{Results}

In this section, we present findings on user editing behavior, long-term interaction patterns with the recommender, misalignment between user-perceived and AI-summarized interest, and a case study comparing user self-written summaries between the formative and post-surveys.

\subsection{Interest Editing}

\begin{figure}[!htbp]
\centering
  \begin{subfigure}[b]{0.4\columnwidth}
    \includegraphics[width=\linewidth]{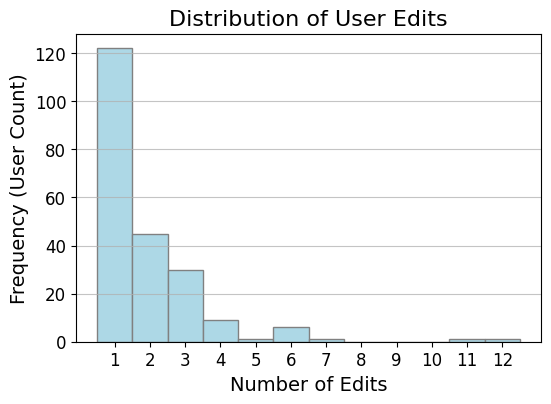}
    \caption{User editing activity distribution.}
    \label{fig:edit-dist}
  \end{subfigure}
  \begin{subfigure}[b]{0.55\columnwidth}
\includegraphics[width=\linewidth]{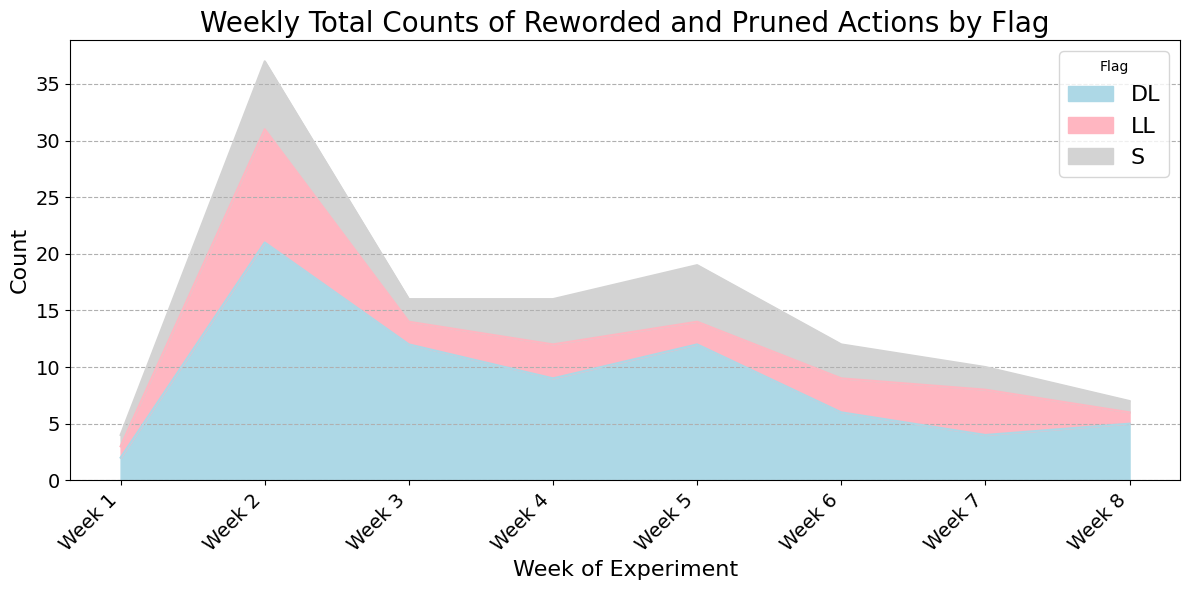}
    \caption{User weekly editing behavior change.}
    \label{fig:sub-weekly-edit}
    \end{subfigure}
  \caption{Distribution of user editing count and chronological change by different interest types.}
\end{figure}

Out of the 1,775 active user participants who visited the movie recommender during our eight-week experiment, 216 interacted with the interest summary interface by editing the Gen-AI-generated summary at least once. On average, each interacting user edited the summary 1.89 times, with a maximum of 12 edits. Detailed editing distribution can be found in Fig. \ref{fig:edit-dist}.

We further examined users' editing behavior within each of the three interest sections at different granularity levels, including both summary and sentence changes. To clearly quantify their modification effort, we applied text embedding cosine similarity on each AI-generated/user-edited summary pairs and categorized it into one of the three levels: \textit{Retained, Reworded}, and \textit{Pruned}, with definitions and examples shown in Table \ref{tab:sentence_editing_examples}. We also provide three summary-level editing examples in Appendix Fig.~\ref{fig:prune-example}.

\begin{table}[!htbp]
    \centering
    \begin{tabularx}{\textwidth}{@{} >{\raggedright\arraybackslash}p{4cm} >{\raggedright\arraybackslash}X >{\raggedright\arraybackslash}X @{}}
        \toprule
        \textbf{Editing Type} & \textbf{AI-Generated} & \textbf{User-Edited} \\
        \midrule

        \textbf{Retained} \newline \textit{Semantic similarity over 95\%} & 
        Movies that feature Michael Caine often evoke a sense of dislike. & 
        Movies that feature Michael Caine often evoke a sense of dislike. \\
        \addlinespace 
        \textbf{Reworded} \newline \textit{Semantic similarity between 60\% and 95\%} & 
        Movies starring Michael J. Fox, Michael Keaton, or Michael Moore are generally not favored. & 
        Movies starring \chdeleted{Michael J. Fox,} \chdeleted{Michael Keaton, or} Michael Moore are generally not favored. \\
        \addlinespace
        \textbf{Pruned} \newline \textit{Semantic similarity below 60\%} & 
        Movies that showcase strong female characters, such as Sarah Michelle Gellar, are often viewed unfavorably &
        \chadded[]{Reality TV is garbage} \\
        \bottomrule
    \end{tabularx}
    \caption{Sentence-level user editing types and corresponding examples.}
    \label{tab:sentence_editing_examples}
\end{table}

\begin{figure}[!htbp]
\centering
  \begin{subfigure}[b]{0.45\columnwidth}
    \includegraphics[width=\linewidth]{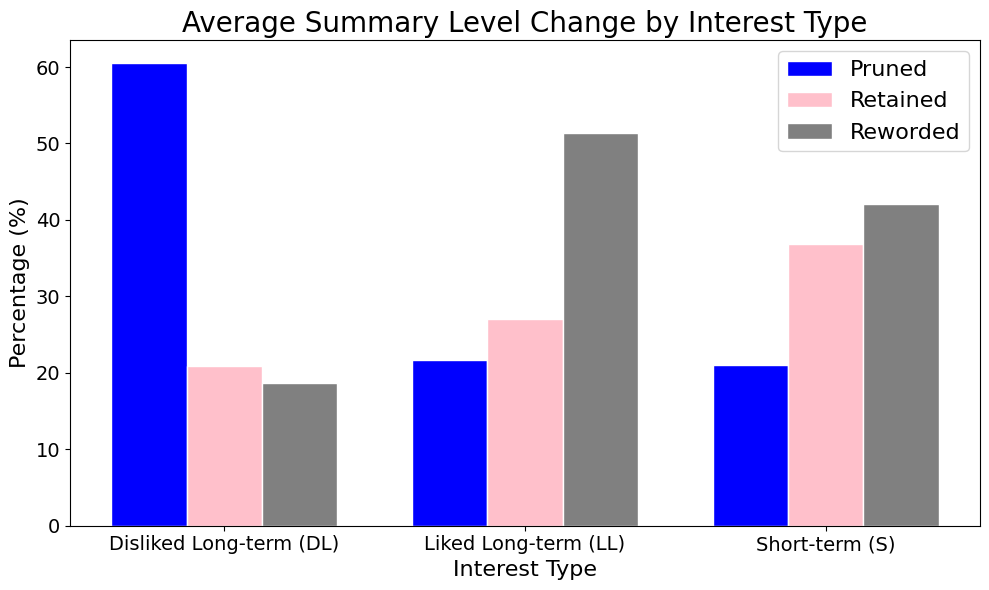}
    \caption{Summary-level interest editing}
    \label{fig:sub-sentence-editing}
  \end{subfigure}
  \begin{subfigure}[b]{0.45\columnwidth}
    \includegraphics[width=\linewidth]{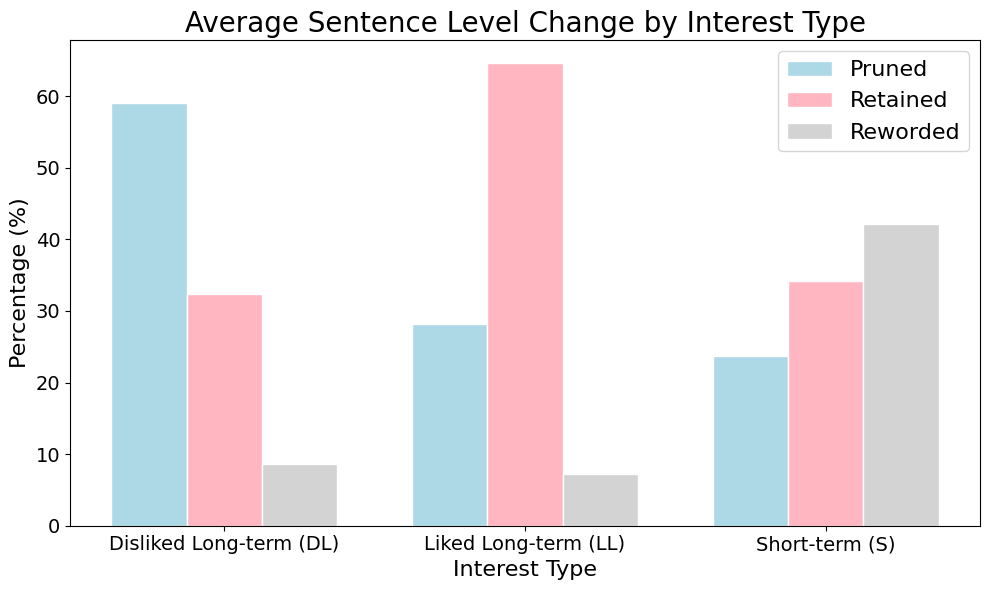}
    \caption{Sentence-level interest editing}
    \label{fig:sub-summary-editing}
  \end{subfigure}
  \caption{Distribution of user editing level by different interest types.}
  \label{fig:editing-level-plots}
\end{figure}

As shown in Fig.~\ref{fig:editing-level-plots}, over half of AI-generated summary and sentence in the \emph{disliked long-term interest} section were pruned by users, whereas nearly two-thirds in the \emph{liked long-term interest} section were retained in sentence level. Users frequently reworded, rather than deleted, short-term interest descriptions, suggesting that AI captured short-term trends relatively well, while users exerted stronger judgment over how these recent interests should be framed. 

We also examined whether user interest descriptions stabilize over time when supported by Gen-AI-assisted reflection. The time-series analysis in Fig.~\ref{fig:sub-weekly-edit} revealed that rewording and pruning activity peaked in Week~2 and gradually declined thereafter, indicating convergence between AI-generated and user-perceived interests along the chronological axis.

\subsection{Impact on Human-Recommender Interaction}

Beyond direct editing behavior, we examined whether the new interest-understanding interface influenced long-term user–recommender interaction patterns. Because we tried to distribute the interface to as many qualified participants as possible if they were willing to try out the new interface, we did not design a specific control group. Instead, we selected the next best approach by splitting the 1,775 qualified participants into three groups: those who collaborated with Gen-AI and edited their interest summary at least twice ($N = 96$), those who interacted with the interface by editing their interest summary once ($N = 120$) and those who only read and reflected on their interest but did not edit ($N = 1,559$). Since people who previously engaged more frequently with the system could bring in selection bias and more likely to engage with the new interface (see pre-experiment log analysis in Appedix table \ref{tab:pre_exp_log_pvals}), we used Analysis of Covariance (ANCOVA)~\cite{frigon1993analysis}, controlling for different group size and each user's baseline behavior during the eight weeks prior to the experiment (the same time of the experiment window), to isolate the effect of different group of users interacting with the new interface. The effect size for ANCOVA is represented by $\eta^2$, with small effect between 0.01 and 0.06, medium effect between 0.06 to 0.14, and large effect over 0.14.

\begin{figure}[!htbp]
\centering
  \includegraphics[width=0.48\columnwidth]{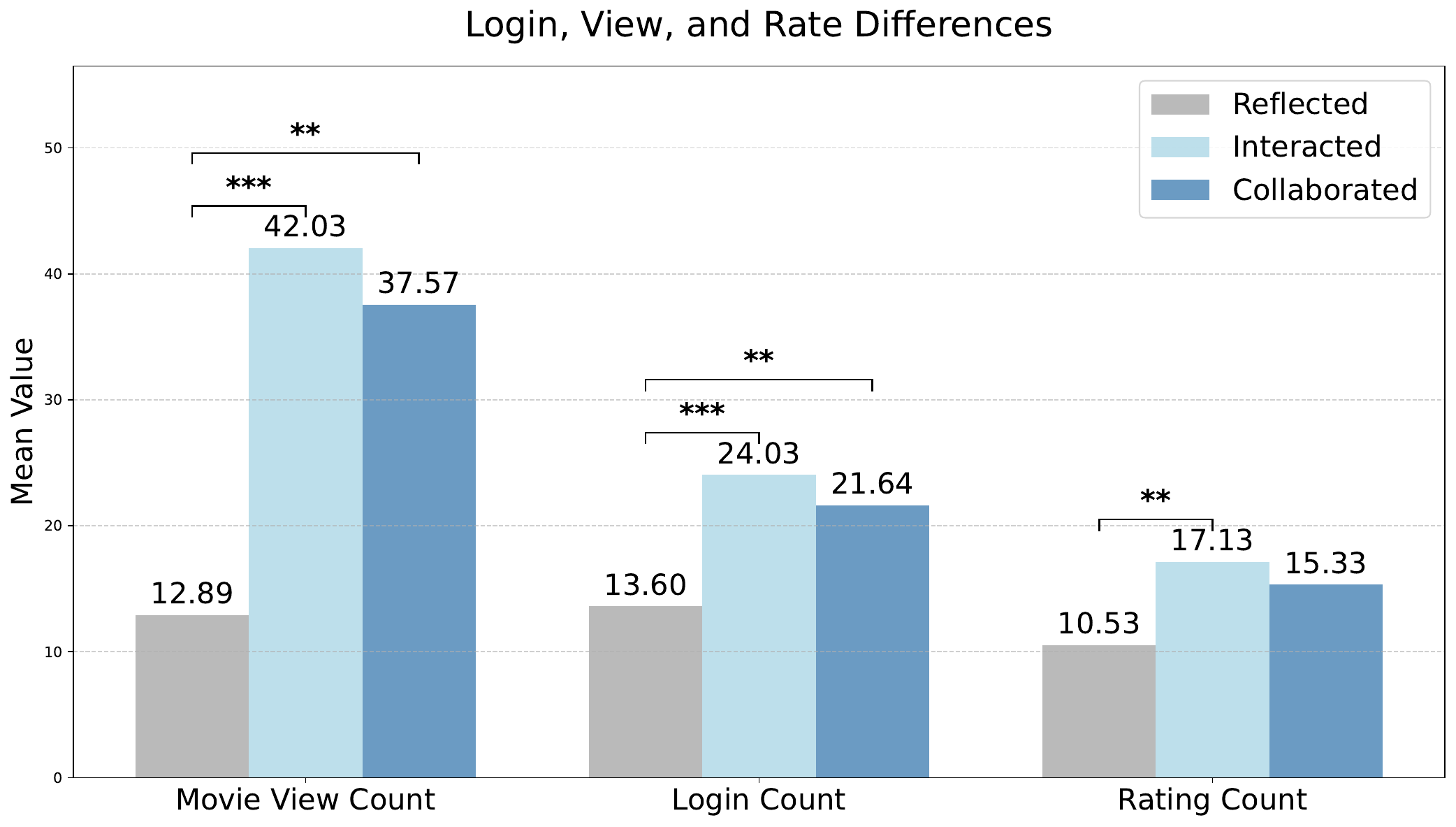}
  \includegraphics[width=0.48\columnwidth]{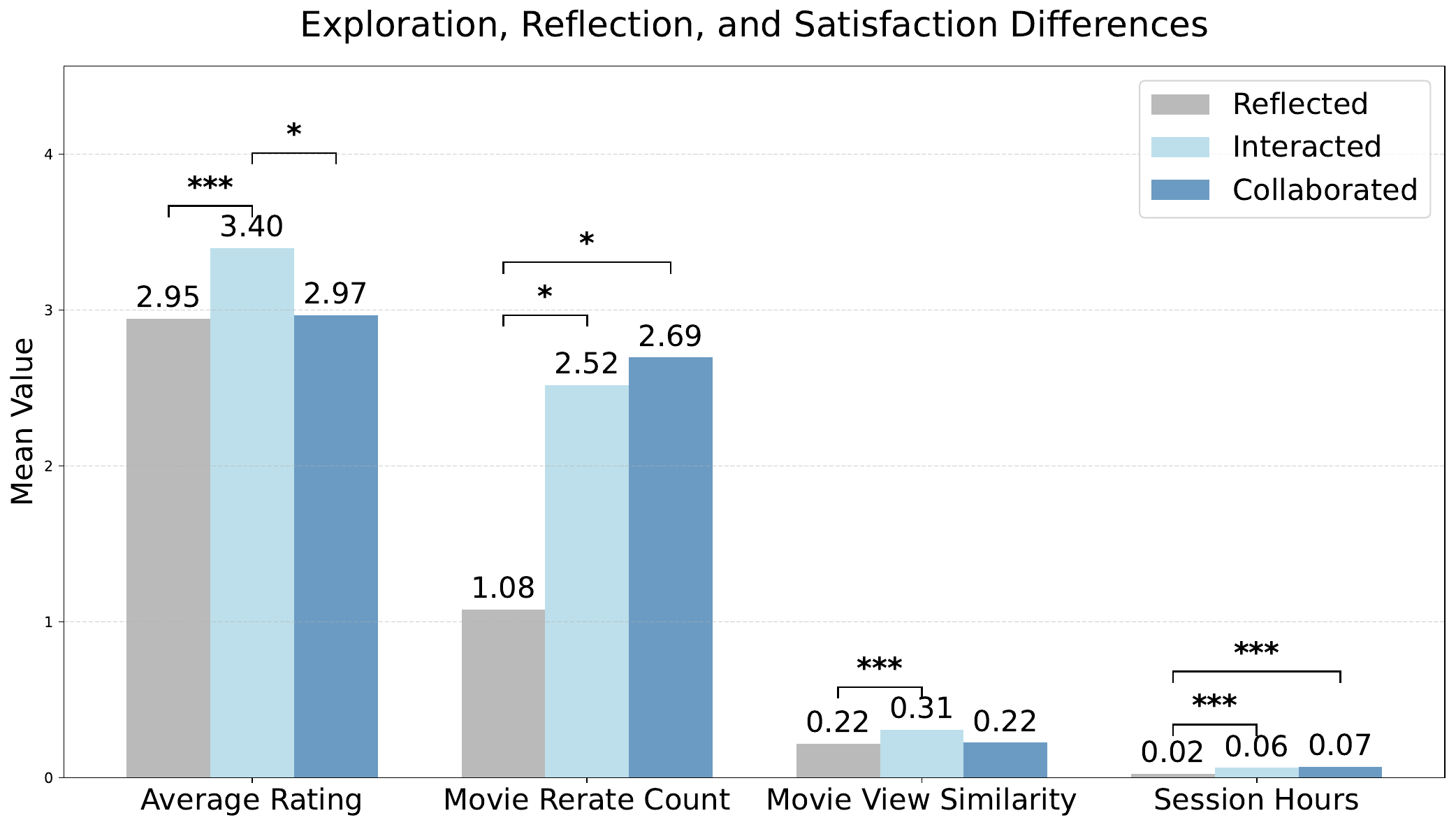}
  \caption{Analysis of Covariance on recommender interaction logs for three groups of participants: Reflected ($N_{edit}=0$), Interacted ($N_{edit}=1$), and Collaborated ($N_{edit} \geq2$). Astericks stands for adjusted p-value for pairwise tukey's HSD comparison: * for p < 0.05, ** for p < 0.01, and *** for p < 0.001.}~\label{fig:ancova-analysis}
\end{figure}

As shown in Fig.~\ref{fig:ancova-analysis}, user behavior differed significantly across the three groups categorized by editing effort. In terms of engagement, both the \textit{Interacted} and \textit{Collaborated} groups were substantially more active than the \textit{Reflected} group. Specifically, the \textit{Interacted} group viewed 226\% more movies ($\eta^2=0.021, p=0.000$) and rated 63\% more movies ($\eta^2=0.004,p=0.001$) than their non-interacting counterparts. Both \textit{Interacted} and \textit{Collaborated} groups also logged in more ($\eta^2=0.001, p_i=0.000, p_c=0.002$) and stayed longer ($\eta^2=0.001, p_i=0.000, p_c=0.000$) compared to the reflected group. Furthermore, interaction appeared to encourage reflection, as both the \textit{Interacted} and \textit{Collaborated} groups re-rated significantly more movies they had previously scored differently ($\eta^2=0.006, p_i=0.0034, p_c=0.0034$), with the \textit{Collaborated} group showed the largest increase at 149\% more re-rates than the \textit{Reflected} group. Interestingly, we observed diversity drop for the \textit{Interacted} group in user exploration, since it demonstrated significantly higher movie view similarity than the \textit{Reflected} group ($\eta^2=0.003,p=0.001$), while the \textit{Collaborated} group view diversity remained unchanged. Finally, for user satisfaction, the average rating score from \textit{Interacted} group was significantly higher than that from both \textit{Reflected} and \textit{Collaborated} groups ($\eta^2=0.005, p_r=0.001, p_c=0.047$).

\subsection{Gap Between User-Perceived and AI-Captured Interests}

\begin{figure}[!htbp]
\centering
  \includegraphics[width=\columnwidth]{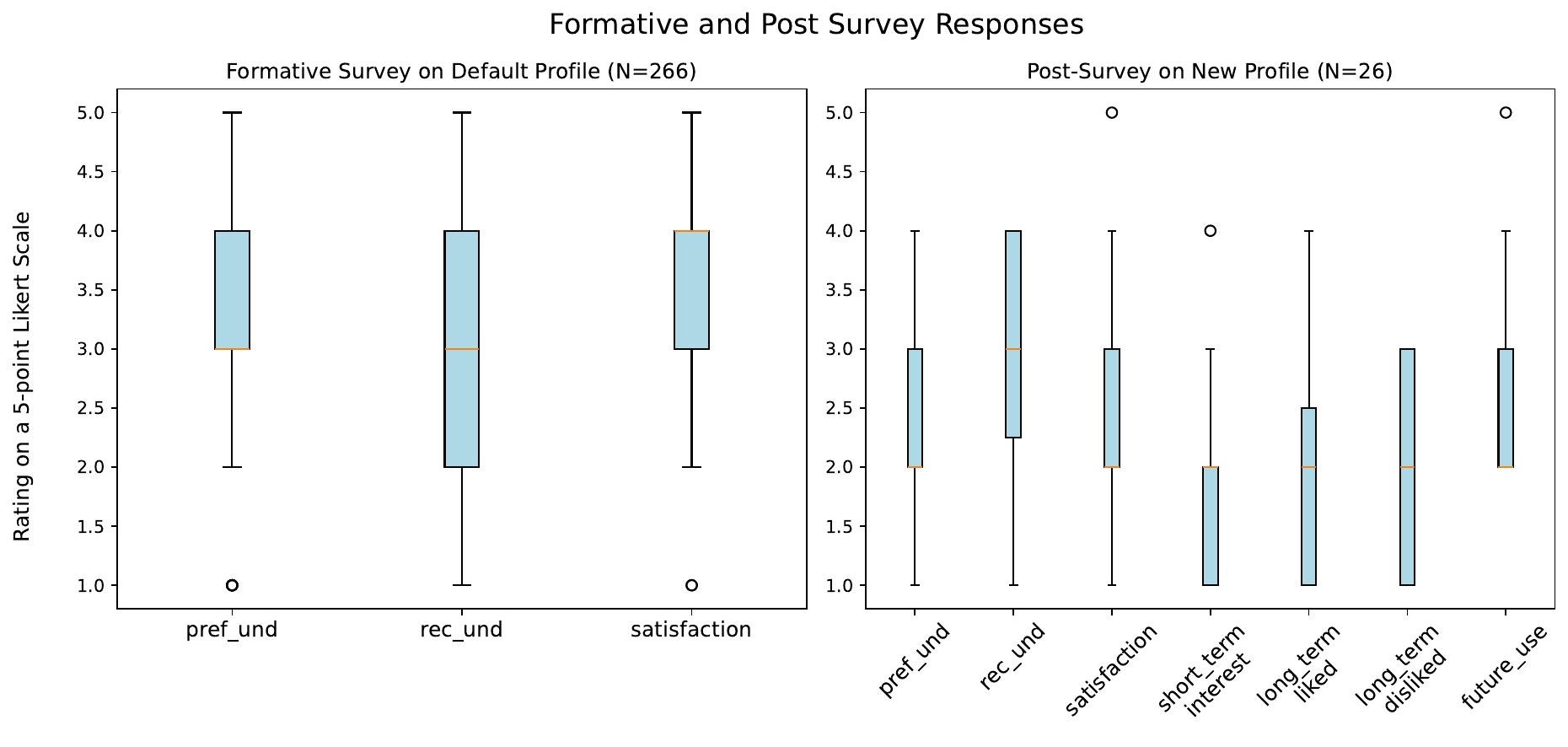}
  \caption{Mean and standard deviation of quantitative survey responses from the formative and post-surveys. Left: User opinions on the default MovieLens profile interface. Right: Responses from users who engaged with the new interface. }~\label{fig:pre-post-survey-stats}
\end{figure}

Post-survey quantitative responses revealed a notable gap between users’ self-perceived interests and how those interests were represented by the system. As demonstrated in the right plot of Fig. \ref{fig:pre-post-survey-stats}, participants reported limited preference understanding from the self-portrait interfaces, and the AI-generated interest summary offered no obvious improvements in recommendation comprehension. Overall satisfaction remained moderate, with participants perceiving limited future utility.

Qualitative coding of the final open-ended question provided additional insight into this misalignment. Many participants felt the initial Gen-AI-generated summary failed to reflect their preferences accurately. For example, one user noted that they watched movies with others and rated them afterward. This co-viewing behavior introduced ratings that did not represent their personal tastes, thereby distorting the AI’s interpretation. Others questioned the AI’s reasoning ability, suggesting it ``overfit'' to sparse data, making overly specific inferences. Mild inaccuracies were also noted, such as unfamiliar directors appearing in summaries. Some users doubted whether highly eclectic or multidimensional interests could ever be captured, while a few expressed general skepticism toward AI personalization, describing the summary as ``AI-slob'' or dismissing the model as ``a global averaging calculator.''

Meanwhile, several participants suggested potential improvements. One proposed that the profile might be most useful for newcomers rather than long-term movie fans who already know their preferences. Another envisioned combining quantitative tag-cluster summaries with qualitative narratives. These comments highlight that users are active interpreters of algorithmic output, seeking both meaning and control.

\subsection{Case Study: Changes in User-Written Self-Portraits}

To better consolidate qualitative insights, we conducted a case study comparing self-written interest descriptions from the formative and post-surveys for four participants. A notable trend was that post-survey statements tended to be more specific and articulate. For example, one user shifted from a broad reflection-\textit{``I think I go through phases of watching different genres, with films in that genre watched back to back''}—to a more precise account: \textit{``I prefer to watch a range of films depending on mood; mainstream comedy or suspense as long as it’s well acted, or I might follow a rabbit hole down a particular foreign director’s oeuvre.''} Another participant moved from a general wish for visualization—\textit{``If I had a graph, or something visual, that would place my ratings in context, it would be nicer''}—to a direct expression of taste: \textit{``I love to search for good horror movies, but I also enjoy hybrid genres and often like movies that are written and/or shot well.''} Similarly, others refined their descriptions, shifting from \textit{``My taste includes young adult, dystopian, supernatural, fantasy, romance, romantic comedy and science fiction''} to emphasizing \textit{``supernatural romance, fantasy, and movies… that require focus in an effort to not miss any major clues or twists.''}

This qualitative shift suggests that interacting with the human-AI collaborated reflection profile may help users express their preferences with greater clarity and detail. However, the degree of change varied; one participant, for example, moved from a broad reflective account to simply referencing specific films: \textit{``Recently hated Ichi the Killer, but historically loved World of Kanako.''} 

\section{Discussion}

As an exploratory online trial to investigate how a human-AI collaborated user profile can help movie recommender users gain better awareness of their own interests, our study reveals several key insights and opportunities for future recommender design and practice.

In response to \emph{\textbf{RQ2}}, our findings show the interface enhances user awareness not by being perfectly accurate, but by serving as a catalyst for self-reflection and interest update within the system. The key insight is that the process of reviewing and editing the AI’s summary is a valuable sense-making activity. This was evident in our case study, where users articulated their tastes with greater specificity after using the profile, and in qualitative feedback where users identified nuanced reasons for inaccuracies, such as co-viewing habits. Besides, we discovered another key value of the interface: users were more willing to edit summaries they disliked compared to those they liked. This behavior is particularly important because negative feedback, while crucial for robust recommendation modeling, is often scarce in practical setting \cite{wang2023learning,liu2025pone,pan2023understanding}. Consequently, the interface can function as an effective tool for collecting explicit user preferences that will benefit future personalization and model training.

Behaviorally, users who interacted and collaborated with AI to edit their interest summaries were more likely to re-rate past movies than the reflection-only group, resulting in a more active refinement of their recommendation preference. However, we have no clear evidence to explain why those who edited their interest only for once demonstrated higher average rating than those who collaborated with Gen-AI and edited more than once (Fig.~\ref{fig:ancova-analysis}). Since the main purpose of website is to provide movie recommendation, one of our assumption is that a single, effective user correction may be optimal for maximizing satisfaction and leaving them more time to explore the recommender, while repeated edits may offer diminishing returns and could signal to the user that the system is struggling to learn their preferences, leading to a less positive experience. Future work can further investigate the real cause behind.

Addressing \emph{\textbf{RQ3}}, we found this critiquing process significantly influenced long-term user-recommender interaction patterns, although with relatively small effect size. Over the eight-week experiment, users who edited the profile showed significant increases in engagement (more logins, views, and ratings), exploration (more diverse movie views) and reflection (more re-ratings) compared to non-interacting users, even after controlling for baseline activity and potential bias. This suggests that empowering users to regularly intervene and co-author their profiles translates into deeper and more varied interaction with the system. Also, since our experiment interface was separated from the regular recommender interaction flow, we speculate about future interfaces that can further benefit user experience from tighter integration. Potential directions include prompting users to re-rate movies when detected significant interest shift from editing, or integrating user-edited interest into recommendation pipeline to refine the items being recommended.

These findings lead to a crucial question for system design: how good does an AI summary have to be to be useful? Our study indicates that the summary does not need to be perfect. In fact, there deserves to be a type of inaccuracy that can serve as a plausible and thought-provoking starting point. With a good design to balance the degree of the discrepancy between the user's self-perception and AI-generated summary, we are producing not a failure but a feature that drives engagement.  This misalignment often stems from concrete data issues like outdated ratings or shared accounts, which opens opportunities to prompt users to provide valuable corrective feedback.

Still, a notable challenge we observed is not just trust in AI, but user's reaction to the modality of the interaction. A declarative, static summary can feel like the AI is "bullying" the user with a simplistic label, leading to resistance. To foster collaboration, systems need a more tactical approach. Instead of presenting a definitive statement, the AI should invite a nuanced conversation, guiding users to elaborate on their preferences in a more interactive and exploratory manner. Moving forward, these insights point toward designing human-in-the-loop systems that prioritize interactive co-creation and critique. The opportunity lies in creating dynamic, conversational systems built on natural language profiles that adapt to preference shifts through direct user edits, as envisioned by \citeauthor{10.1145/3477495.3531873} \cite{10.1145/3477495.3531873}. Indeed, recent work has already demonstrated that such profiles can improve the offline accuracy of movie and book recommendations \cite{zhou2024languagebaseduserprofilesrecommendation,gao2024end}, supporting the significant potential of this user-driven approach to personalization.

\section{Practical and Societal Implications}

Our work points to a practical shift in the design of user-facing recommender systems. Rather than pursuing a perfectly accurate summary that can be perceived as declarative and alienating, designers should aim to create interfaces that invite a dialogue. The goal is a "productively wrong" starting point that leverages misalignment to prompt user reflection and consequently engage with both AI and the system to correct outdated history data. As our findings demonstrate, this collaborative process not only improves the system's understanding but also directly translates into deeper user engagement and reflection.

On a broader level, co-creative systems foster greater algorithmic agency. By making the AI’s inferences transparent and editable, these tools empower users to actively manage their digital identities rather than passively accepting algorithmic labels. This provides a tangible mechanism for users to push back against filter bubbles and steer their own discovery. As AI-generated summaries of people become more common, our work advocates for an ethical design pattern where individuals are not just subjects of AI analysis, but active collaborators with the right to review, contest, and refine their own digital portraits.

\section{Conclusion}

In this study, we introduced a human-AI collaborative interface for co-creating a user's interest profile in a movie recommender system. Through a longitudinal online trial, we demonstrated that while discrepancies persist between user-perceived and system-captured interests, the process of critiquing the AI-generated summary enhances user self-awareness and directly leads to more engaged and reflective interaction with the recommender. Our findings highlight the value of designing AI not as a perfect expert, but as a collaborative partner that scaffolds user reflection and control, paving the way for more transparent and trustworthy recommender experiences.

\bibliographystyle{ACM-Reference-Format}
\bibliography{sample-base}


\begin{thebibliography}{38}


\ifx \showCODEN    \undefined \def \showCODEN     #1{\unskip}     \fi
\ifx \showISBNx    \undefined \def \showISBNx     #1{\unskip}     \fi
\ifx \showISBNxiii \undefined \def \showISBNxiii  #1{\unskip}     \fi
\ifx \showISSN     \undefined \def \showISSN      #1{\unskip}     \fi
\ifx \showLCCN     \undefined \def \showLCCN      #1{\unskip}     \fi
\ifx \shownote     \undefined \def \shownote      #1{#1}          \fi
\ifx \showarticletitle \undefined \def \showarticletitle #1{#1}   \fi
\ifx \showURL      \undefined \def \showURL       {\relax}        \fi
\providecommand\bibfield[2]{#2}
\providecommand\bibinfo[2]{#2}
\providecommand\natexlab[1]{#1}
\providecommand\showeprint[2][]{arXiv:#2}

\bibitem[Balog et~al\mbox{.}(2019)]%
        {10.1145/3331184.3331211}
\bibfield{author}{\bibinfo{person}{Krisztian Balog}, \bibinfo{person}{Filip Radlinski}, {and} \bibinfo{person}{Shushan Arakelyan}.} \bibinfo{year}{2019}\natexlab{}.
\newblock \showarticletitle{Transparent, Scrutable and Explainable User Models for Personalized Recommendation}. In \bibinfo{booktitle}{\emph{Proceedings of the 42nd International ACM SIGIR Conference on Research and Development in Information Retrieval}} (Paris, France) \emph{(\bibinfo{series}{SIGIR'19})}. \bibinfo{publisher}{Association for Computing Machinery}, \bibinfo{address}{New York, NY, USA}, \bibinfo{pages}{265–274}.
\newblock
\showISBNx{9781450361729}
\href{https://doi.org/10.1145/3331184.3331211}{doi:\nolinkurl{10.1145/3331184.3331211}}


\bibitem[Bang and Song(2025)]%
        {bang2025llmbaseduserprofilemanagement}
\bibfield{author}{\bibinfo{person}{Seunghwan Bang} {and} \bibinfo{person}{Hwanjun Song}.} \bibinfo{year}{2025}\natexlab{}.
\newblock \bibinfo{title}{LLM-based User Profile Management for Recommender System}.
\newblock
\showeprint[arxiv]{2502.14541}~[cs.CL]
\urldef\tempurl%
\url{https://arxiv.org/abs/2502.14541}
\showURL{%
\tempurl}


\bibitem[Chen et~al\mbox{.}(2023)]%
        {chen2023bias}
\bibfield{author}{\bibinfo{person}{Jiawei Chen}, \bibinfo{person}{Hande Dong}, \bibinfo{person}{Xiang Wang}, \bibinfo{person}{Fuli Feng}, \bibinfo{person}{Meng Wang}, {and} \bibinfo{person}{Xiangnan He}.} \bibinfo{year}{2023}\natexlab{}.
\newblock \showarticletitle{Bias and debias in recommender system: A survey and future directions}.
\newblock \bibinfo{journal}{\emph{ACM Transactions on Information Systems}} \bibinfo{volume}{41}, \bibinfo{number}{3} (\bibinfo{year}{2023}), \bibinfo{pages}{1--39}.
\newblock


\bibitem[Chen et~al\mbox{.}(2015)]%
        {chen2015recommender}
\bibfield{author}{\bibinfo{person}{Li Chen}, \bibinfo{person}{Guanliang Chen}, {and} \bibinfo{person}{Feng Wang}.} \bibinfo{year}{2015}\natexlab{}.
\newblock \showarticletitle{Recommender systems based on user reviews: the state of the art}.
\newblock \bibinfo{journal}{\emph{User Modeling and User-Adapted Interaction}} \bibinfo{volume}{25}, \bibinfo{number}{2} (\bibinfo{year}{2015}), \bibinfo{pages}{99--154}.
\newblock


\bibitem[Devlin et~al\mbox{.}(2019)]%
        {Devlin_Chang_Lee_Toutanova_2019}
\bibfield{author}{\bibinfo{person}{Jacob Devlin}, \bibinfo{person}{Ming-Wei Chang}, \bibinfo{person}{Kenton Lee}, {and} \bibinfo{person}{Kristina Toutanova}.} \bibinfo{year}{2019}\natexlab{}.
\newblock \showarticletitle{BERT: Pre-training of Deep Bidirectional Transformers for Language Understanding}. In \bibinfo{booktitle}{\emph{Proceedings of the 2019 Conference of the North American Chapter of the Association for Computational Linguistics: Human Language Technologies, Volume 1 (Long and Short Papers)}}, \bibfield{editor}{\bibinfo{person}{Jill Burstein}, \bibinfo{person}{Christy Doran}, {and} \bibinfo{person}{Thamar Solorio}} (Eds.). \bibinfo{publisher}{Association for Computational Linguistics}, \bibinfo{address}{Minneapolis, Minnesota}, \bibinfo{pages}{4171–4186}.
\newblock
\href{https://doi.org/10.18653/v1/N19-1423}{doi:\nolinkurl{10.18653/v1/N19-1423}}


\bibitem[Dhillon et~al\mbox{.}(2024)]%
        {10.1145/3613904.3642134}
\bibfield{author}{\bibinfo{person}{Paramveer~S. Dhillon}, \bibinfo{person}{Somayeh Molaei}, \bibinfo{person}{Jiaqi Li}, \bibinfo{person}{Maximilian Golub}, \bibinfo{person}{Shaochun Zheng}, {and} \bibinfo{person}{Lionel~Peter Robert}.} \bibinfo{year}{2024}\natexlab{}.
\newblock \showarticletitle{Shaping Human-AI Collaboration: Varied Scaffolding Levels in Co-writing with Language Models}. In \bibinfo{booktitle}{\emph{Proceedings of the 2024 CHI Conference on Human Factors in Computing Systems}} (Honolulu, HI, USA) \emph{(\bibinfo{series}{CHI '24})}. \bibinfo{publisher}{Association for Computing Machinery}, \bibinfo{address}{New York, NY, USA}, Article \bibinfo{articleno}{1044}, \bibinfo{numpages}{18}~pages.
\newblock
\showISBNx{9798400703300}
\href{https://doi.org/10.1145/3613904.3642134}{doi:\nolinkurl{10.1145/3613904.3642134}}


\bibitem[Eiband et~al\mbox{.}(2021)]%
        {eiband2021support}
\bibfield{author}{\bibinfo{person}{Malin Eiband}, \bibinfo{person}{Daniel Buschek}, {and} \bibinfo{person}{Heinrich Hussmann}.} \bibinfo{year}{2021}\natexlab{}.
\newblock \showarticletitle{How to support users in understanding intelligent systems? Structuring the discussion}. In \bibinfo{booktitle}{\emph{Proceedings of the 26th International Conference on Intelligent User Interfaces}}. \bibinfo{pages}{120--132}.
\newblock


\bibitem[Frigon and Laurencelle(1993)]%
        {frigon1993analysis}
\bibfield{author}{\bibinfo{person}{Jean-Yves Frigon} {and} \bibinfo{person}{Louis Laurencelle}.} \bibinfo{year}{1993}\natexlab{}.
\newblock \showarticletitle{Analysis of covariance: A proposed algorithm}.
\newblock \bibinfo{journal}{\emph{Educational and Psychological Measurement}} \bibinfo{volume}{53}, \bibinfo{number}{1} (\bibinfo{year}{1993}), \bibinfo{pages}{1--18}.
\newblock


\bibitem[Gao et~al\mbox{.}(2024)]%
        {gao2024end}
\bibfield{author}{\bibinfo{person}{Zhaolin Gao}, \bibinfo{person}{Joyce Zhou}, \bibinfo{person}{Yijia Dai}, {and} \bibinfo{person}{Thorsten Joachims}.} \bibinfo{year}{2024}\natexlab{}.
\newblock \showarticletitle{End-to-end Training for Recommendation with Language-based User Profiles}.
\newblock \bibinfo{journal}{\emph{arXiv preprint arXiv:2410.18870}} (\bibinfo{year}{2024}).
\newblock


\bibitem[Grootendorst(2022)]%
        {grootendorst2022bertopicneuraltopicmodeling}
\bibfield{author}{\bibinfo{person}{Maarten Grootendorst}.} \bibinfo{year}{2022}\natexlab{}.
\newblock \bibinfo{title}{BERTopic: Neural topic modeling with a class-based TF-IDF procedure}.
\newblock
\showeprint[arxiv]{2203.05794}~[cs.CL]
\urldef\tempurl%
\url{https://arxiv.org/abs/2203.05794}
\showURL{%
\tempurl}


\bibitem[Guo et~al\mbox{.}(2024)]%
        {10.1145/3613905.3651042}
\bibfield{author}{\bibinfo{person}{Jiajing Guo}, \bibinfo{person}{Vikram Mohanty}, \bibinfo{person}{Jorge~H Piazentin~Ono}, \bibinfo{person}{Hongtao Hao}, \bibinfo{person}{Liang Gou}, {and} \bibinfo{person}{Liu Ren}.} \bibinfo{year}{2024}\natexlab{}.
\newblock \showarticletitle{Investigating Interaction Modes and User Agency in Human-LLM Collaboration for Domain-Specific Data Analysis}. In \bibinfo{booktitle}{\emph{Extended Abstracts of the CHI Conference on Human Factors in Computing Systems}} (Honolulu, HI, USA) \emph{(\bibinfo{series}{CHI EA '24})}. \bibinfo{publisher}{Association for Computing Machinery}, \bibinfo{address}{New York, NY, USA}, Article \bibinfo{articleno}{203}, \bibinfo{numpages}{9}~pages.
\newblock
\showISBNx{9798400703317}
\href{https://doi.org/10.1145/3613905.3651042}{doi:\nolinkurl{10.1145/3613905.3651042}}


\bibitem[He et~al\mbox{.}(2023)]%
        {he2023survey}
\bibfield{author}{\bibinfo{person}{Zhicheng He}, \bibinfo{person}{Weiwen Liu}, \bibinfo{person}{Wei Guo}, \bibinfo{person}{Jiarui Qin}, \bibinfo{person}{Yingxue Zhang}, \bibinfo{person}{Yaochen Hu}, {and} \bibinfo{person}{Ruiming Tang}.} \bibinfo{year}{2023}\natexlab{}.
\newblock \showarticletitle{A survey on user behavior modeling in recommender systems}.
\newblock \bibinfo{journal}{\emph{arXiv preprint arXiv:2302.11087}} (\bibinfo{year}{2023}).
\newblock


\bibitem[Hill et~al\mbox{.}(2025)]%
        {Hill_Goo_Agarwal_2025}
\bibfield{author}{\bibinfo{person}{Alessandro Hill}, \bibinfo{person}{Kalen Goo}, {and} \bibinfo{person}{Puneet Agarwal}.} \bibinfo{year}{2025}\natexlab{}.
\newblock \showarticletitle{Recommending the right academic programs: an interest mining approach using BERTopic}.
\newblock \bibinfo{journal}{\emph{Data Mining and Knowledge Discovery}} \bibinfo{volume}{39}, \bibinfo{number}{3} (\bibinfo{date}{March} \bibinfo{year}{2025}), \bibinfo{pages}{20}.
\newblock
\showISSN{1573-756X}
\href{https://doi.org/10.1007/s10618-024-01087-y}{doi:\nolinkurl{10.1007/s10618-024-01087-y}}


\bibitem[Jannach et~al\mbox{.}(2021)]%
        {jannach2021survey}
\bibfield{author}{\bibinfo{person}{Dietmar Jannach}, \bibinfo{person}{Ahtsham Manzoor}, \bibinfo{person}{Wanling Cai}, {and} \bibinfo{person}{Li Chen}.} \bibinfo{year}{2021}\natexlab{}.
\newblock \showarticletitle{A survey on conversational recommender systems}.
\newblock \bibinfo{journal}{\emph{ACM Computing Surveys (CSUR)}} \bibinfo{volume}{54}, \bibinfo{number}{5} (\bibinfo{year}{2021}), \bibinfo{pages}{1--36}.
\newblock


\bibitem[Jesse et~al\mbox{.}(2023)]%
        {jesse2023intra}
\bibfield{author}{\bibinfo{person}{Mathias Jesse}, \bibinfo{person}{Christine Bauer}, {and} \bibinfo{person}{Dietmar Jannach}.} \bibinfo{year}{2023}\natexlab{}.
\newblock \showarticletitle{Intra-list similarity and human diversity perceptions of recommendations: the details matter: M. Jesse et al.}
\newblock \bibinfo{journal}{\emph{User Modeling and User-Adapted Interaction}} \bibinfo{volume}{33}, \bibinfo{number}{4} (\bibinfo{year}{2023}), \bibinfo{pages}{769--802}.
\newblock


\bibitem[Kim et~al\mbox{.}(2024a)]%
        {kim-etal-2024-meganno}
\bibfield{author}{\bibinfo{person}{Hannah Kim}, \bibinfo{person}{Kushan Mitra}, \bibinfo{person}{Rafael Li~Chen}, \bibinfo{person}{Sajjadur Rahman}, {and} \bibinfo{person}{Dan Zhang}.} \bibinfo{year}{2024}\natexlab{a}.
\newblock \showarticletitle{{MEGA}nno+: A Human-{LLM} Collaborative Annotation System}. In \bibinfo{booktitle}{\emph{Proceedings of the 18th Conference of the European Chapter of the Association for Computational Linguistics: System Demonstrations}}, \bibfield{editor}{\bibinfo{person}{Nikolaos Aletras} {and} \bibinfo{person}{Orphee De~Clercq}} (Eds.). \bibinfo{publisher}{Association for Computational Linguistics}, \bibinfo{address}{St. Julians, Malta}, \bibinfo{pages}{168--176}.
\newblock
\urldef\tempurl%
\url{https://aclanthology.org/2024.eacl-demo.18/}
\showURL{%
\tempurl}


\bibitem[Kim et~al\mbox{.}(2024b)]%
        {10.1145/3613904.3642693}
\bibfield{author}{\bibinfo{person}{Taewan Kim}, \bibinfo{person}{Donghoon Shin}, \bibinfo{person}{Young-Ho Kim}, {and} \bibinfo{person}{Hwajung Hong}.} \bibinfo{year}{2024}\natexlab{b}.
\newblock \showarticletitle{DiaryMate: Understanding User Perceptions and Experience in Human-AI Collaboration for Personal Journaling}. In \bibinfo{booktitle}{\emph{Proceedings of the 2024 CHI Conference on Human Factors in Computing Systems}} (Honolulu, HI, USA) \emph{(\bibinfo{series}{CHI '24})}. \bibinfo{publisher}{Association for Computing Machinery}, \bibinfo{address}{New York, NY, USA}, Article \bibinfo{articleno}{1046}, \bibinfo{numpages}{15}~pages.
\newblock
\showISBNx{9798400703300}
\href{https://doi.org/10.1145/3613904.3642693}{doi:\nolinkurl{10.1145/3613904.3642693}}


\bibitem[Knijnenburg et~al\mbox{.}(2012)]%
        {knijnenburg2012explaining}
\bibfield{author}{\bibinfo{person}{Bart~P Knijnenburg}, \bibinfo{person}{Martijn~C Willemsen}, \bibinfo{person}{Zeno Gantner}, \bibinfo{person}{Hakan Soncu}, {and} \bibinfo{person}{Chris Newell}.} \bibinfo{year}{2012}\natexlab{}.
\newblock \showarticletitle{Explaining the user experience of recommender systems}.
\newblock \bibinfo{journal}{\emph{User modeling and user-adapted interaction}}  \bibinfo{volume}{22} (\bibinfo{year}{2012}), \bibinfo{pages}{441--504}.
\newblock


\bibitem[Kronhardt et~al\mbox{.}(2024)]%
        {10.1145/3701571.3703373}
\bibfield{author}{\bibinfo{person}{Kirill Kronhardt}, \bibinfo{person}{Sebastian Hoffmann}, \bibinfo{person}{Fabian Adelt}, {and} \bibinfo{person}{Jens Gerken}.} \bibinfo{year}{2024}\natexlab{}.
\newblock \showarticletitle{PERSON\AE{}R - Transparency Enhancing Tool for LLM-Generated User Personas from Live Website Visits}. In \bibinfo{booktitle}{\emph{Proceedings of the International Conference on Mobile and Ubiquitous Multimedia}} \emph{(\bibinfo{series}{MUM '24})}. \bibinfo{publisher}{Association for Computing Machinery}, \bibinfo{address}{New York, NY, USA}, \bibinfo{pages}{527–531}.
\newblock
\showISBNx{9798400712838}
\href{https://doi.org/10.1145/3701571.3703373}{doi:\nolinkurl{10.1145/3701571.3703373}}


\bibitem[Lacerda and Ziviani(2013)]%
        {10.1145/2433396.2433492}
\bibfield{author}{\bibinfo{person}{Anisio Lacerda} {and} \bibinfo{person}{Nivio Ziviani}.} \bibinfo{year}{2013}\natexlab{}.
\newblock \showarticletitle{Building user profiles to improve user experience in recommender systems}. In \bibinfo{booktitle}{\emph{Proceedings of the Sixth ACM International Conference on Web Search and Data Mining}} (Rome, Italy) \emph{(\bibinfo{series}{WSDM '13})}. \bibinfo{publisher}{Association for Computing Machinery}, \bibinfo{address}{New York, NY, USA}, \bibinfo{pages}{759–764}.
\newblock
\showISBNx{9781450318693}
\href{https://doi.org/10.1145/2433396.2433492}{doi:\nolinkurl{10.1145/2433396.2433492}}


\bibitem[Liu et~al\mbox{.}(2025)]%
        {liu2025pone}
\bibfield{author}{\bibinfo{person}{Ziyang Liu}, \bibinfo{person}{Chaokun Wang}, \bibinfo{person}{Shuwen Zheng}, \bibinfo{person}{Cheng Wu}, \bibinfo{person}{Kai Zheng}, \bibinfo{person}{Yang Song}, {and} \bibinfo{person}{Na Mou}.} \bibinfo{year}{2025}\natexlab{}.
\newblock \showarticletitle{Pone-GNN: integrating positive and negative feedback in graph neural networks for recommender systems}.
\newblock \bibinfo{journal}{\emph{ACM Transactions on Recommender Systems}} \bibinfo{volume}{3}, \bibinfo{number}{2} (\bibinfo{year}{2025}), \bibinfo{pages}{1--23}.
\newblock


\bibitem[Lu et~al\mbox{.}(2023)]%
        {lu2023user}
\bibfield{author}{\bibinfo{person}{Hongyu Lu}, \bibinfo{person}{Weizhi Ma}, \bibinfo{person}{Yifan Wang}, \bibinfo{person}{Min Zhang}, \bibinfo{person}{Xiang Wang}, \bibinfo{person}{Yiqun Liu}, \bibinfo{person}{Tat-Seng Chua}, {and} \bibinfo{person}{Shaoping Ma}.} \bibinfo{year}{2023}\natexlab{}.
\newblock \showarticletitle{User perception of recommendation explanation: Are your explanations what users need?}
\newblock \bibinfo{journal}{\emph{ACM Transactions on Information Systems}} \bibinfo{volume}{41}, \bibinfo{number}{2} (\bibinfo{year}{2023}), \bibinfo{pages}{1--31}.
\newblock


\bibitem[Maslowska et~al\mbox{.}(2022)]%
        {maslowska2022role}
\bibfield{author}{\bibinfo{person}{Ewa Maslowska}, \bibinfo{person}{Edward~C Malthouse}, {and} \bibinfo{person}{Linda~D Hollebeek}.} \bibinfo{year}{2022}\natexlab{}.
\newblock \showarticletitle{The role of recommender systems in fostering consumers' long-term platform engagement}.
\newblock \bibinfo{journal}{\emph{Journal of Service Management}} \bibinfo{volume}{33}, \bibinfo{number}{4/5} (\bibinfo{year}{2022}), \bibinfo{pages}{721--732}.
\newblock


\bibitem[Pan et~al\mbox{.}(2023)]%
        {pan2023understanding}
\bibfield{author}{\bibinfo{person}{Yunzhu Pan}, \bibinfo{person}{Chen Gao}, \bibinfo{person}{Jianxin Chang}, \bibinfo{person}{Yanan Niu}, \bibinfo{person}{Yang Song}, \bibinfo{person}{Kun Gai}, \bibinfo{person}{Depeng Jin}, {and} \bibinfo{person}{Yong Li}.} \bibinfo{year}{2023}\natexlab{}.
\newblock \showarticletitle{Understanding and modeling passive-negative feedback for short-video sequential recommendation}. In \bibinfo{booktitle}{\emph{Proceedings of the 17th ACM conference on recommender systems}}. \bibinfo{pages}{540--550}.
\newblock


\bibitem[Pu et~al\mbox{.}(2012)]%
        {pu2012evaluating}
\bibfield{author}{\bibinfo{person}{Pearl Pu}, \bibinfo{person}{Li Chen}, {and} \bibinfo{person}{Rong Hu}.} \bibinfo{year}{2012}\natexlab{}.
\newblock \showarticletitle{Evaluating recommender systems from the user’s perspective: survey of the state of the art}.
\newblock \bibinfo{journal}{\emph{User Modeling and User-Adapted Interaction}}  \bibinfo{volume}{22} (\bibinfo{year}{2012}), \bibinfo{pages}{317--355}.
\newblock


\bibitem[Radlinski et~al\mbox{.}(2022)]%
        {10.1145/3477495.3531873}
\bibfield{author}{\bibinfo{person}{Filip Radlinski}, \bibinfo{person}{Krisztian Balog}, \bibinfo{person}{Fernando Diaz}, \bibinfo{person}{Lucas Dixon}, {and} \bibinfo{person}{Ben Wedin}.} \bibinfo{year}{2022}\natexlab{}.
\newblock \showarticletitle{On Natural Language User Profiles for Transparent and Scrutable Recommendation}. In \bibinfo{booktitle}{\emph{Proceedings of the 45th International ACM SIGIR Conference on Research and Development in Information Retrieval}} (Madrid, Spain) \emph{(\bibinfo{series}{SIGIR '22})}. \bibinfo{publisher}{Association for Computing Machinery}, \bibinfo{address}{New York, NY, USA}, \bibinfo{pages}{2863–2874}.
\newblock
\showISBNx{9781450387323}
\href{https://doi.org/10.1145/3477495.3531873}{doi:\nolinkurl{10.1145/3477495.3531873}}


\bibitem[Rana et~al\mbox{.}(2024)]%
        {10.1145/3597499}
\bibfield{author}{\bibinfo{person}{Arpit Rana}, \bibinfo{person}{Scott Sanner}, \bibinfo{person}{Mohamed~Reda Bouadjenek}, \bibinfo{person}{Ronald Di~Carlantonio}, {and} \bibinfo{person}{Gary Farmaner}.} \bibinfo{year}{2024}\natexlab{}.
\newblock \showarticletitle{User Experience and the Role of Personalization in Critiquing-Based Conversational Recommendation}.
\newblock \bibinfo{journal}{\emph{ACM Trans. Web}} \bibinfo{volume}{18}, \bibinfo{number}{4}, Article \bibinfo{articleno}{43} (\bibinfo{date}{Oct.} \bibinfo{year}{2024}), \bibinfo{numpages}{21}~pages.
\newblock
\showISSN{1559-1131}
\href{https://doi.org/10.1145/3597499}{doi:\nolinkurl{10.1145/3597499}}


\bibitem[Sidji et~al\mbox{.}(2024)]%
        {10.1145/3677081}
\bibfield{author}{\bibinfo{person}{Matthew Sidji}, \bibinfo{person}{Wally Smith}, {and} \bibinfo{person}{Melissa~J. Rogerson}.} \bibinfo{year}{2024}\natexlab{}.
\newblock \showarticletitle{Human-AI Collaboration in Cooperative Games: A Study of Playing Codenames with an LLM Assistant}.
\newblock \bibinfo{journal}{\emph{Proc. ACM Hum.-Comput. Interact.}} \bibinfo{volume}{8}, \bibinfo{number}{CHI PLAY}, Article \bibinfo{articleno}{316} (\bibinfo{date}{Oct.} \bibinfo{year}{2024}), \bibinfo{numpages}{25}~pages.
\newblock
\href{https://doi.org/10.1145/3677081}{doi:\nolinkurl{10.1145/3677081}}


\bibitem[Tintarev and Masthoff(2010)]%
        {tintarev2010designing}
\bibfield{author}{\bibinfo{person}{Nava Tintarev} {and} \bibinfo{person}{Judith Masthoff}.} \bibinfo{year}{2010}\natexlab{}.
\newblock \showarticletitle{Designing and evaluating explanations for recommender systems}.
\newblock In \bibinfo{booktitle}{\emph{Recommender systems handbook}}. \bibinfo{publisher}{Springer}, \bibinfo{pages}{479--510}.
\newblock


\bibitem[Tsai and Brusilovsky(2021)]%
        {tsai2021effects}
\bibfield{author}{\bibinfo{person}{Chun-Hua Tsai} {and} \bibinfo{person}{Peter Brusilovsky}.} \bibinfo{year}{2021}\natexlab{}.
\newblock \showarticletitle{The effects of controllability and explainability in a social recommender system}.
\newblock \bibinfo{journal}{\emph{User Modeling and User-Adapted Interaction}}  \bibinfo{volume}{31} (\bibinfo{year}{2021}), \bibinfo{pages}{591--627}.
\newblock


\bibitem[Vig et~al\mbox{.}(2011)]%
        {vig2011navigating}
\bibfield{author}{\bibinfo{person}{Jesse Vig}, \bibinfo{person}{Shilad Sen}, {and} \bibinfo{person}{John Riedl}.} \bibinfo{year}{2011}\natexlab{}.
\newblock \showarticletitle{Navigating the tag genome}. In \bibinfo{booktitle}{\emph{Proceedings of the 16th international conference on Intelligent user interfaces}}. \bibinfo{pages}{93--102}.
\newblock


\bibitem[Wang et~al\mbox{.}(2020)]%
        {10.1145/3334480.3381069}
\bibfield{author}{\bibinfo{person}{Dakuo Wang}, \bibinfo{person}{Elizabeth Churchill}, \bibinfo{person}{Pattie Maes}, \bibinfo{person}{Xiangmin Fan}, \bibinfo{person}{Ben Shneiderman}, \bibinfo{person}{Yuanchun Shi}, {and} \bibinfo{person}{Qianying Wang}.} \bibinfo{year}{2020}\natexlab{}.
\newblock \showarticletitle{From Human-Human Collaboration to Human-AI Collaboration: Designing AI Systems That Can Work Together with People}. In \bibinfo{booktitle}{\emph{Extended Abstracts of the 2020 CHI Conference on Human Factors in Computing Systems}} (Honolulu, HI, USA) \emph{(\bibinfo{series}{CHI EA '20})}. \bibinfo{publisher}{Association for Computing Machinery}, \bibinfo{address}{New York, NY, USA}, \bibinfo{pages}{1–6}.
\newblock
\showISBNx{9781450368193}
\href{https://doi.org/10.1145/3334480.3381069}{doi:\nolinkurl{10.1145/3334480.3381069}}


\bibitem[Wang et~al\mbox{.}(2023)]%
        {wang2023learning}
\bibfield{author}{\bibinfo{person}{Yueqi Wang}, \bibinfo{person}{Yoni Halpern}, \bibinfo{person}{Shuo Chang}, \bibinfo{person}{Jingchen Feng}, \bibinfo{person}{Elaine~Ya Le}, \bibinfo{person}{Longfei Li}, \bibinfo{person}{Xujian Liang}, \bibinfo{person}{Min-Cheng Huang}, \bibinfo{person}{Shane Li}, \bibinfo{person}{Alex Beutel}, {et~al\mbox{.}}} \bibinfo{year}{2023}\natexlab{}.
\newblock \showarticletitle{Learning from negative user feedback and measuring responsiveness for sequential recommenders}. In \bibinfo{booktitle}{\emph{Proceedings of the 17th ACM Conference on Recommender Systems}}. \bibinfo{pages}{1049--1053}.
\newblock


\bibitem[Xue et~al\mbox{.}(2023)]%
        {xue2023prefrec}
\bibfield{author}{\bibinfo{person}{Wanqi Xue}, \bibinfo{person}{Qingpeng Cai}, \bibinfo{person}{Zhenghai Xue}, \bibinfo{person}{Shuo Sun}, \bibinfo{person}{Shuchang Liu}, \bibinfo{person}{Dong Zheng}, \bibinfo{person}{Peng Jiang}, \bibinfo{person}{Kun Gai}, {and} \bibinfo{person}{Bo An}.} \bibinfo{year}{2023}\natexlab{}.
\newblock \showarticletitle{Prefrec: Recommender systems with human preferences for reinforcing long-term user engagement}. In \bibinfo{booktitle}{\emph{Proceedings of the 29th ACM SIGKDD Conference on Knowledge Discovery and Data Mining}}. \bibinfo{pages}{2874--2884}.
\newblock


\bibitem[Yang et~al\mbox{.}(2015)]%
        {Yang_Song_Ji_2015}
\bibfield{author}{\bibinfo{person}{Ping Yang}, \bibinfo{person}{Yan Song}, {and} \bibinfo{person}{Yang Ji}.} \bibinfo{year}{2015}\natexlab{}.
\newblock \showarticletitle{Tag-based user interest discovery though keywords extraction in social network}. In \bibinfo{booktitle}{\emph{Big Data Computing and Communications}}, \bibfield{editor}{\bibinfo{person}{Yu~Wang}, \bibinfo{person}{Hui Xiong}, \bibinfo{person}{Shlomo Argamon}, \bibinfo{person}{XiangYang Li}, {and} \bibinfo{person}{JianZhong Li}} (Eds.). \bibinfo{publisher}{Springer International Publishing}, \bibinfo{address}{Cham}, \bibinfo{pages}{363–372}.
\newblock
\showISBNx{9783319220475}
\href{https://doi.org/10.1007/978-3-319-22047-5_29}{doi:\nolinkurl{10.1007/978-3-319-22047-5_29}}


\bibitem[Yu et~al\mbox{.}(2024)]%
        {fi16070254}
\bibfield{author}{\bibinfo{person}{Rui Yu}, \bibinfo{person}{Sooyeon Lee}, \bibinfo{person}{Jingyi Xie}, \bibinfo{person}{Syed~Masum Billah}, {and} \bibinfo{person}{John~M. Carroll}.} \bibinfo{year}{2024}\natexlab{}.
\newblock \showarticletitle{Human–AI Collaboration for Remote Sighted Assistance: Perspectives from the LLM Era}.
\newblock \bibinfo{journal}{\emph{Future Internet}} \bibinfo{volume}{16}, \bibinfo{number}{7} (\bibinfo{year}{2024}).
\newblock
\showISSN{1999-5903}
\href{https://doi.org/10.3390/fi16070254}{doi:\nolinkurl{10.3390/fi16070254}}


\bibitem[Zhao et~al\mbox{.}(2018)]%
        {zhao2018explicit}
\bibfield{author}{\bibinfo{person}{Qian Zhao}, \bibinfo{person}{F~Maxwell Harper}, \bibinfo{person}{Gediminas Adomavicius}, {and} \bibinfo{person}{Joseph~A Konstan}.} \bibinfo{year}{2018}\natexlab{}.
\newblock \showarticletitle{Explicit or implicit feedback? Engagement or satisfaction? A field experiment on machine-learning-based recommender systems}. In \bibinfo{booktitle}{\emph{Proceedings of the 33rd Annual ACM symposium on applied computing}}. \bibinfo{pages}{1331--1340}.
\newblock


\bibitem[Zhou et~al\mbox{.}(2024)]%
        {zhou2024languagebaseduserprofilesrecommendation}
\bibfield{author}{\bibinfo{person}{Joyce Zhou}, \bibinfo{person}{Yijia Dai}, {and} \bibinfo{person}{Thorsten Joachims}.} \bibinfo{year}{2024}\natexlab{}.
\newblock \bibinfo{title}{Language-Based User Profiles for Recommendation}.
\newblock
\showeprint[arxiv]{2402.15623}~[cs.CL]
\urldef\tempurl%
\url{https://arxiv.org/abs/2402.15623}
\showURL{%
\tempurl}


\end{thebibliography}

\appendix

\section*{Appendices}
\setcounter{table}{0}
\renewcommand{\thetable}{A\arabic{table}}

\begin{table}[ht]
\centering
\begin{tabular}{p{14cm}}
\toprule
\textbf{Formative Survey Questions} \\
\midrule
Q1. Are you aware of the "about your ratings" page on MovieLens? \\
\addlinespace
Q2. How useful do you find the information on this page for understanding your preferences? \\
\addlinespace
Q3. How easy is it for you to understand why certain movies are recommended to you, based on your rating profile? \\
\addlinespace
Q4. How satisfied are you with your current rating profile? \\
\addlinespace
Q5. Based on your profile, can you write a short paragraph describing your taste in movies? You might include preferred genres, themes, or notable films or plots you enjoy. \\
\addlinespace
Q6. If you could make new changes to your user profile, what new components or features would you wish to see? Please list and describe your ideas if applicable. \\
\bottomrule
\end{tabular}
\caption{Formative Survey Questions}
\label{tab:pre_survey_questions}
\end{table}

\begin{table}[h]
\centering
\begin{tabular}{p{14cm}}
\toprule
\textbf{Post Survey Questions} \\
\midrule
Q1. In the past several weeks, did you check out the "Your Interest Summary" page on MovieLens? (If No, survey ends) \\
\addlinespace
Q2. How useful do you find the information on the "Your Interest Summary" page for understanding your preferences? \\
\addlinespace
Q3. How easy is it for you to understand why certain movies are recommended to you, based on your interest summary profile? \\
\addlinespace
Q4. How satisfied are you with the interest summary profile? \\
\addlinespace
Q5. In the "Self Portrait" section, you have seen three sections of editable text summary. How would you rate the "What you liked recently" summary? \\
\addlinespace
Q6. How would you rate the "Some long-term interests" summary? \\
\addlinespace
Q7. How would you rate the "Looks like you do not enjoy..." summary? \\
\addlinespace
Q8. How often would you use the "Your Interest Summary" page if it became a future feature on the MovieLens website? \\
\addlinespace
Q9. Based on your understanding of your interest, can you write a short paragraph describing your taste in movies again? You might include preferred genres, themes, cast, or notable films or plots you enjoy. \\
\addlinespace
Q10. What else would you like us to know about your experience with the interest summary page? \\
\bottomrule
\end{tabular}
\caption{Post Survey Questions used in study}
\label{tab:post_survey_questions}
\end{table}

\begin{table}[htbp]
\centering
\begin{tabular}{lrrrr}
\toprule
\textrm{Metric} & \textrm{ANOVA} & \textrm{Ref-Int} & \textrm{Ref-Col} & \textrm{Int-Col} \\
\midrule
movie\_view\_count & 0.0884 & 0.1109 & 0.5517 & 0.8332 \\
rating\_count & 0.0003*** & 0.0080** & 0.0123* & 0.9879 \\
login\_count & 0.0000*** & 0.0000*** & 0.0000*** & 0.9705 \\
session\_length & 0.0000*** & 0.0000*** & 0.0000*** & 0.0848 \\
rated\_movie\_div & 0.0497* & 0.0427* & 0.1740 & 0.9880 \\
viewed\_movie\_div & 0.0151* & 0.0125* & 0.1166 & 0.9419 \\
rerate\_total & 0.0000*** & 0.0000*** & 0.0000*** & 0.7711 \\
avg\_rating & 0.1423 & 0.2248 & 0.1458 & 0.9999 \\
\bottomrule
\end{tabular}
\caption{Pre-experiment ANOVA and pairwise p-values on log analysis metrics. Ref stands for reflected, Int stands for interacted, and Col stands for collaborated group. Significance levels are denoted by: * for p < 0.05, ** for p < 0.01, and *** for p < 0.001.}
\label{tab:pre_exp_log_pvals}
\end{table}

\setcounter{figure}{0}
\renewcommand{\thefigure}{A\arabic{figure}}

\begin{figure}[!htbp]
\centering
  \includegraphics[width=\columnwidth]{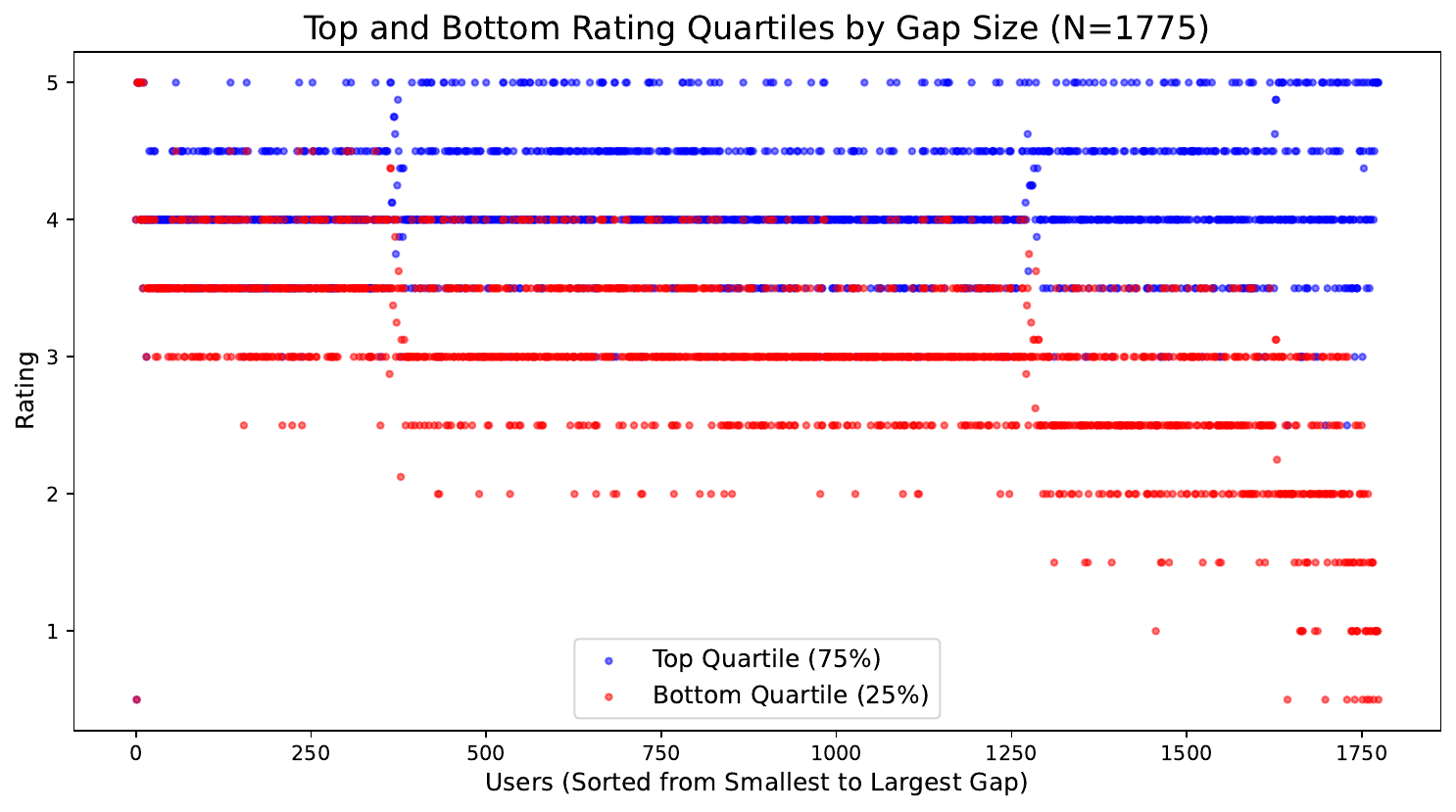}
  \caption{Top and bottom rating quartile cut-off for qualified user long-term liked and disliked summary generation.}~\label{fig:top-bottom-rating-quartiles}
\end{figure}

\begin{figure}[!htbp]
\centering
  \begin{subfigure}[b]{\columnwidth}
    \includegraphics[width=\linewidth]{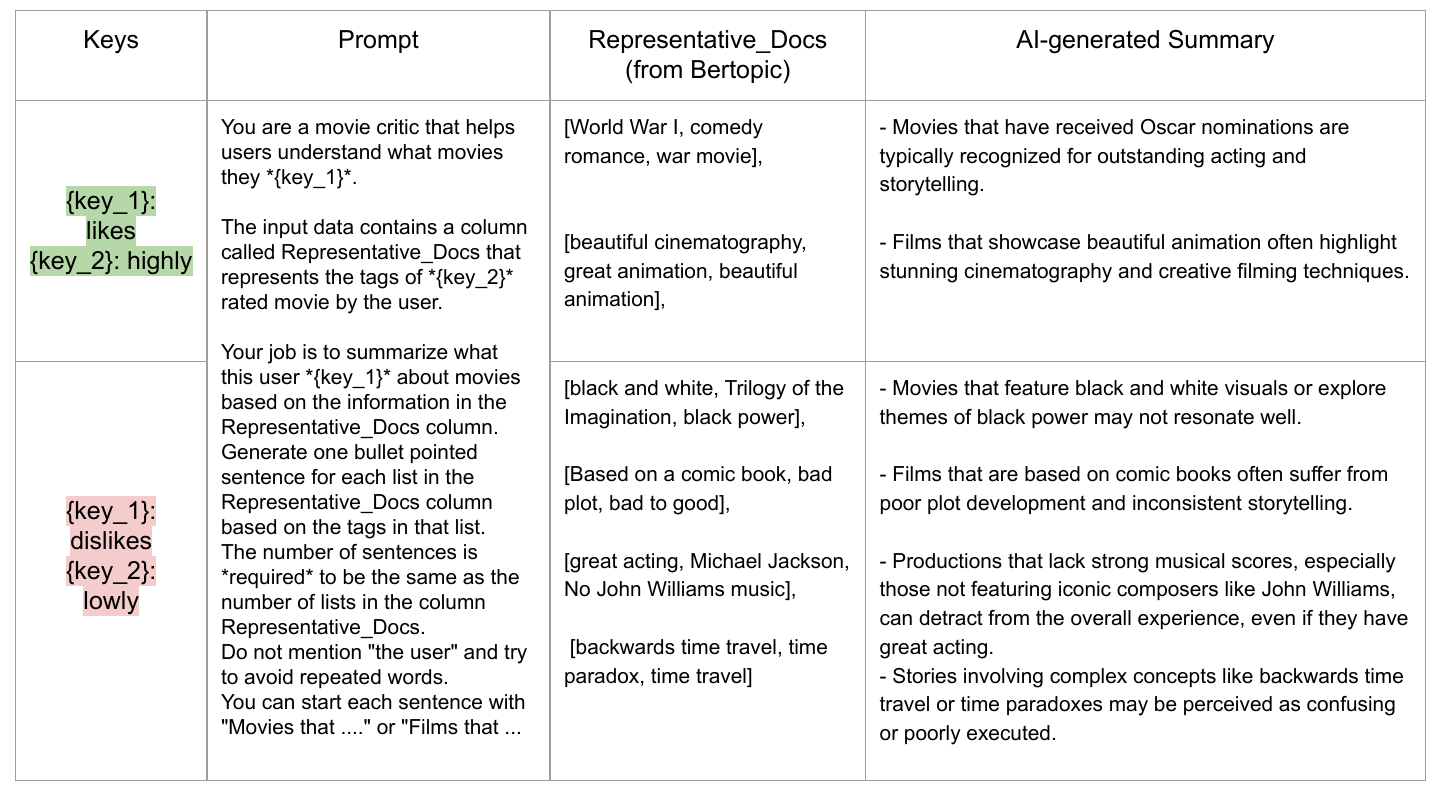}
    \caption{Initial long-term interest prompt and output example for both liked and disliked summary generation.}
    \label{fig:generation_prompt_1}
  \end{subfigure}
  \hfill
  \begin{subfigure}[b]{\columnwidth}
    \includegraphics[width=\linewidth]{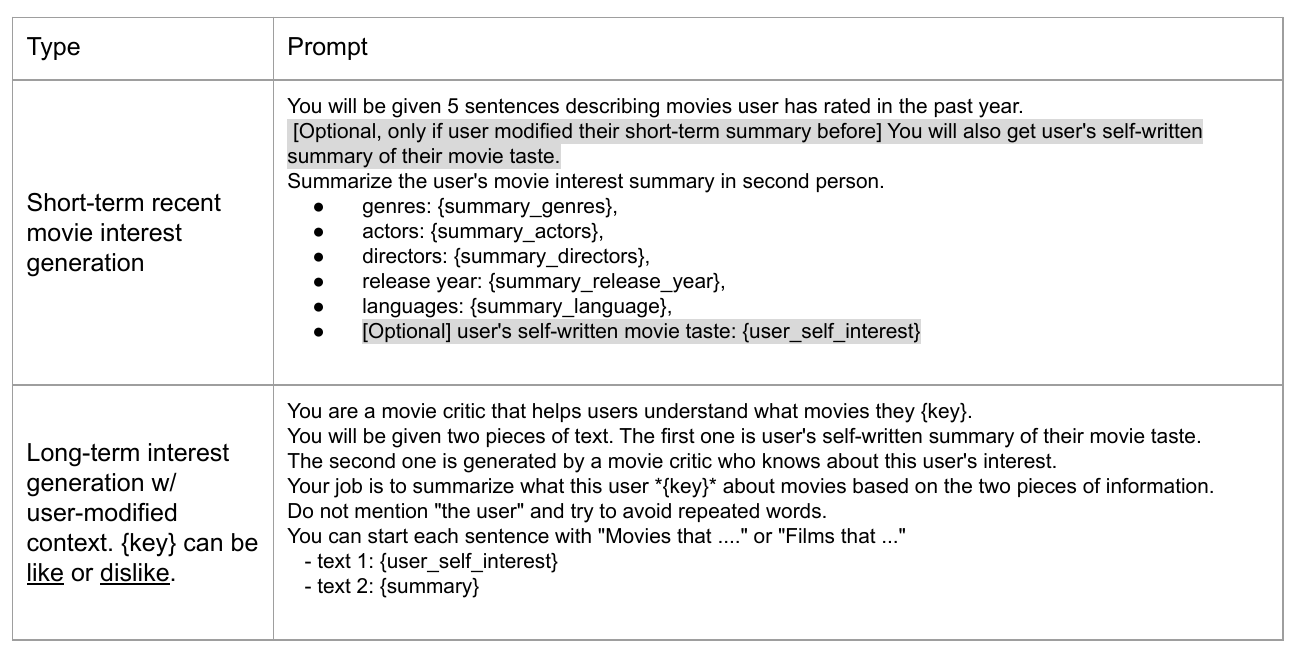}
    \caption{Short-term interest prompt and long-term summary generation examples with user editing context.}
    \label{fig:generation_prompt_2}
  \end{subfigure}
  \caption{User interest prompt examples}
  \label{fig:generation_prompt}
\end{figure}

\begin{figure}[!htbp]
\centering
  \includegraphics[width=\columnwidth]{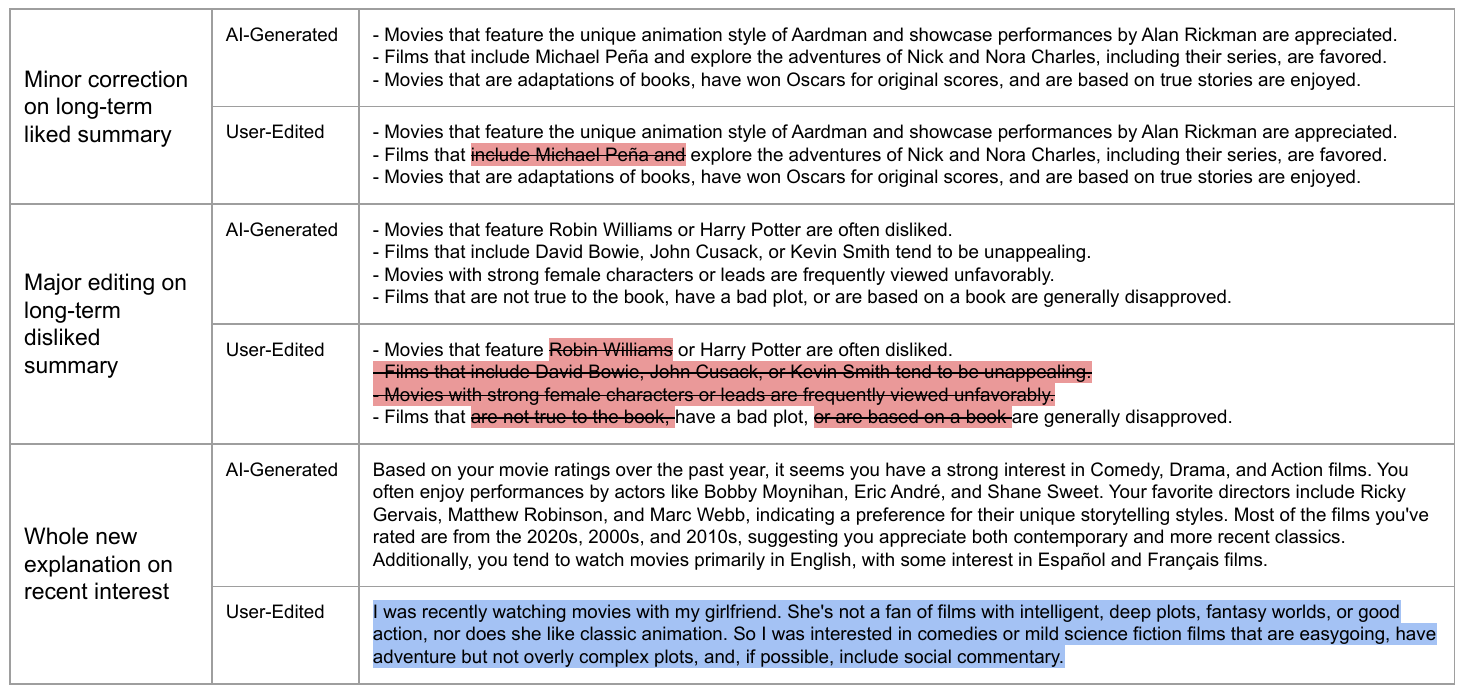}
  \caption{Different summary-level modification examples}~\label{fig:prune-example}
\end{figure}

\end{document}